\documentclass[twocolumn]{aastex631}

\usepackage{natbib}
\usepackage{multirow}
\usepackage{color}
\usepackage{bm}
\usepackage[normalem]{ulem}

\newcommand{\eg}{e.g., }
\newcommand{\ie}{i.e., }
\newcommand{\Msun}{M_{\odot}}

\newcommand{\Mej}{M_{\rm ej}}

\def\gsim{\mathrel{\rlap{\lower 4pt \hbox{\hskip 1pt $\sim$}}\raise 1pt \hbox {$>$}}}
\def\lsim{\mathrel{\rlap{\lower 4pt \hbox{\hskip 1pt $\sim$}}\raise 1pt \hbox {$<$}}}

\shorttitle{Lanthanide Features in Near-infrared Spectra of Kilonovae}
\shortauthors{N. Domoto et al.}

\begin{document}

\title{Lanthanide Features in Near-infrared Spectra of Kilonovae} 

\correspondingauthor{Nanae Domoto}
\email{n.domoto@astr.tohoku.ac.jp}

\author[0000-0002-7415-7954]{Nanae Domoto}
\affiliation{Astronomical Institute, Tohoku University, Aoba, Sendai 980-8578, Japan}

\author[0000-0001-8253-6850]{Masaomi Tanaka}
\affiliation{Astronomical Institute, Tohoku University, Aoba, Sendai 980-8578, Japan}
\affiliation{Division for the Establishment of Frontier Sciences, Organization for Advanced Studies, Tohoku University, Sendai 980-8577, Japan}

\author[0000-0002-5302-073X]{Daiji Kato}
\affiliation{National Institute for Fusion Science, 322-6 Oroshi-cho, Toki 509-5292, Japan}
\affiliation{Interdisciplinary Graduate School of Engineering Sciences, Kyushu University, Kasuga, Fukuoka 816-8580, Japan}

\author[0000-0003-4443-6984]{Kyohei Kawaguchi}
\affiliation{Institute for Cosmic Ray Research, The University of Tokyo, 5-1-5 Kashiwanoha, Kashiwa, Chiba 277-8582, Japan}
\affiliation{Center for Gravitational Physics, Yukawa Institute for Theoretical Physics, Kyoto University, Kyoto 606-8502, Japan}

\author[0000-0002-2502-3730]{Kenta Hotokezaka}
\affiliation{Research Center for the Early Universe, Graduate School of Science, University of Tokyo, Bunkyo, Tokyo 113-0033, Japan}
\affiliation{Kavli IPMU (WPI), UTIAS, The University of Tokyo, Kashiwa, Chiba 277-8583, Japan}

\author[0000-0002-4759-7794]{Shinya Wanajo}
\affiliation{Max-Planck-Institut f\"{u}r Gravitationsphysik (Albert-Einstein-Institut), Am M\"{u}hlenberg 1, D-14476 Potsdam-Golm, Germany}
\affiliation{Interdisciplinary Theoretical and Mathematical Sciences Program (iTHEMS), RIKEN, Wako, Saitama 351-0198, Japan}



\begin{abstract}
The observations of GW170817/AT2017gfo have provided us with evidence that binary neutron star mergers are sites of $r$-process nucleosynthesis.
However, the observed signatures in the spectra of GW170817/AT2017gfo have not been fully decoded especially in the near-infrared (NIR) wavelengths.
In this paper, we investigate the kilonova spectra over the entire wavelength range with the aim of elemental identification.
We systematically calculate the strength of bound-bound transitions by constructing a hybrid line list that is accurate for important strong transitions and complete for weak transitions.
We find that the elements on the left side of the periodic table, such as Ca, Sr, Y, Zr, Ba, La, and Ce, tend to produce prominent absorption lines in the spectra.
This is because such elements have a small number of valence electrons and low-lying energy levels, resulting in strong transitions.
By performing self-consistent radiative transfer simulations for the entire ejecta, we find that La III and Ce III appear in the NIR spectra, which can explain the absorption features at $\lambda\sim 12000$--14000 {\AA} in the spectra of GW170817/AT2017gfo.
The mass fractions of La and Ce are estimated to be $>2\times 10^{-6}$ and $\sim$ (1--100)$\times 10^{-5}$, respectively.
An actinide element Th can also be a source of absorption as the atomic structure is analogous to that of Ce.
However, we show that Th III features are less prominent in the spectra because of the denser energy levels of actinides compared to those of lanthanides.
\end{abstract}



\section{Introduction}
\label{sec:intro}
Binary neutron star (NS) mergers are promising sites of rapid neutron capture nucleosynthesis \citep[$r$-process, e.g.,][]{LS1974, Eichler1989, Meyer1989, Freiburghaus1999, Goriely2011a, Korobkin2012, Wanajo2014}.
Radioactive decay of freshly synthesized nuclei in the ejected neutron-rich material powers electromagnetic emission called a kilonova \citep{LiPaczynski1998, Metzger2010, Roberts2011}.
In 2017, together with the detection of gravitational waves (GW) from a NS merger \citep[GW170817,][]{Abbott2017a}, an electromagnetic counterpart was identified \citep[AT2017gfo,][]{Abbott2017b}.
The observed properties of AT2017gfo in ultraviolet, optical, and near-infrared (NIR) wavelengths are consistent with the theoretical expectation of a kilonova \citep[e.g.,][]{Arcavi2017, Coulter2017, Evans2017, Pian2017, Smartt2017, Utsumi2017, Valenti2017}.
The electromagnetic counterpart has provided us with evidence that NS mergers are sites of $r$-process nucleosynthesis \citep[e.g.,][]{Kasen2017, Perego2017, Shibata2017, Tanaka2017, Kawaguchi2018, Rosswog2018}.

It is important to reveal the abundance pattern in NS merger ejecta.
However, the elemental abundances synthesized in GW170817 are not yet clear.
One of the direct ways of finding the synthesized elements is the identification of absorption lines in photospheric spectra.
\citet{Watson2019} analyzed the observed spectra of GW170817/AT2017gfo at a few days after the merger.
Based on spectral calculations above the photosphere, they found that the absorption features around $\lambda\sim 8000$ {\AA} could be explained by the Sr II lines \citep[see also][]{Gillanders2022}.
\citet{Domoto2021} carried out self-consistent radiative transfer simulations of the entire ejecta, and also showed that Sr II produces strong absorption lines.
\citet{Perego2022} suggested the presence of He as an alternative explanation of the features around $\lambda\sim 8000$ {\AA}, but ultimately concluded that this was unlikely.
While the observed spectra exhibit several features, especially at NIR wavelengths, no other element has yet been identified (see \citealp{Gillanders2021} for a search of Pt and Au).

Synthesized elements can also be identified in the spectra during the nebula phase because of the appearance of emission lines.
In GW170817, the Spitzer space telescope detected the late-time nebular emission at 4.5 $\mu$m and put an upper limit at 3.6 $\mu$m, suggesting distinctive spectral features \citep{Villar2018, Kasliwal2022}.
Recently, \citet{Hotokezaka2021} and \citet{Pognan2022a, Pognan2022b} have initiated work on the nebula phase of a kilonova.
Although conclusive identification of elements has not been made with the spectra, it is suggested that the IR nebula emission in GW170817 can be explained mainly by the lines of Se ($Z=34$) or W ($Z=74$) \citep{Hotokezaka2022}.

One of the issues facing the study of kilonova spectra is a lack of atomic data for heavy elements.
To extract elemental information from the spectra, spectroscopically accurate atomic data are needed.
However, since such atomic data are not complete, especially for heavy elements at NIR wavelengths, we have been able to investigate lines only at the optical wavelengths, $\lambda \lesssim 10000$ {\AA} \citep{Watson2019, Domoto2021}.
Although the incompleteness of the data can be mitigated by theoretical calculations \citep[e.g.,][]{Kasen2013, Tanaka2018, Tanaka2020, Fontes2020, Banerjee2020, Pognan2022b}, such theoretical data require calibration with experimental data for quantitative discussion on spectral features, due to the low accuracy in wavelengths \citep{Gillanders2021}.

In this paper, we propose a new scheme to investigate the spectral features over the whole wavelength range with the aim of elemental identification in kilonova photospheric spectra.
In Section \ref{sec:linelist}, we systematically calculate the strength of bound-bound transitions by means of a simple one-zone model using theoretical atomic data.
By combining atomic data based on theoretical calculations and experiments, we construct a new hybrid line list that is accurate for important strong transitions and complete for weak transitions.
Then, in Section \ref{sec:spectra}, we perform radiative transfer simulations of NS merger ejecta with the new line list.
In Section \ref{sec:discussion}, we discuss the estimated lanthanide abundances in GW170817/AT2017gfo and possibilities of identifying actinide elements.
Finally, we give our conclusions in Section \ref{sec:conclusion}.

\section{Line list}
\label{sec:linelist}
\begin{figure*}[ht]
  \begin{center}
    \begin{tabular}{cc}
    \includegraphics[width=0.48\linewidth]{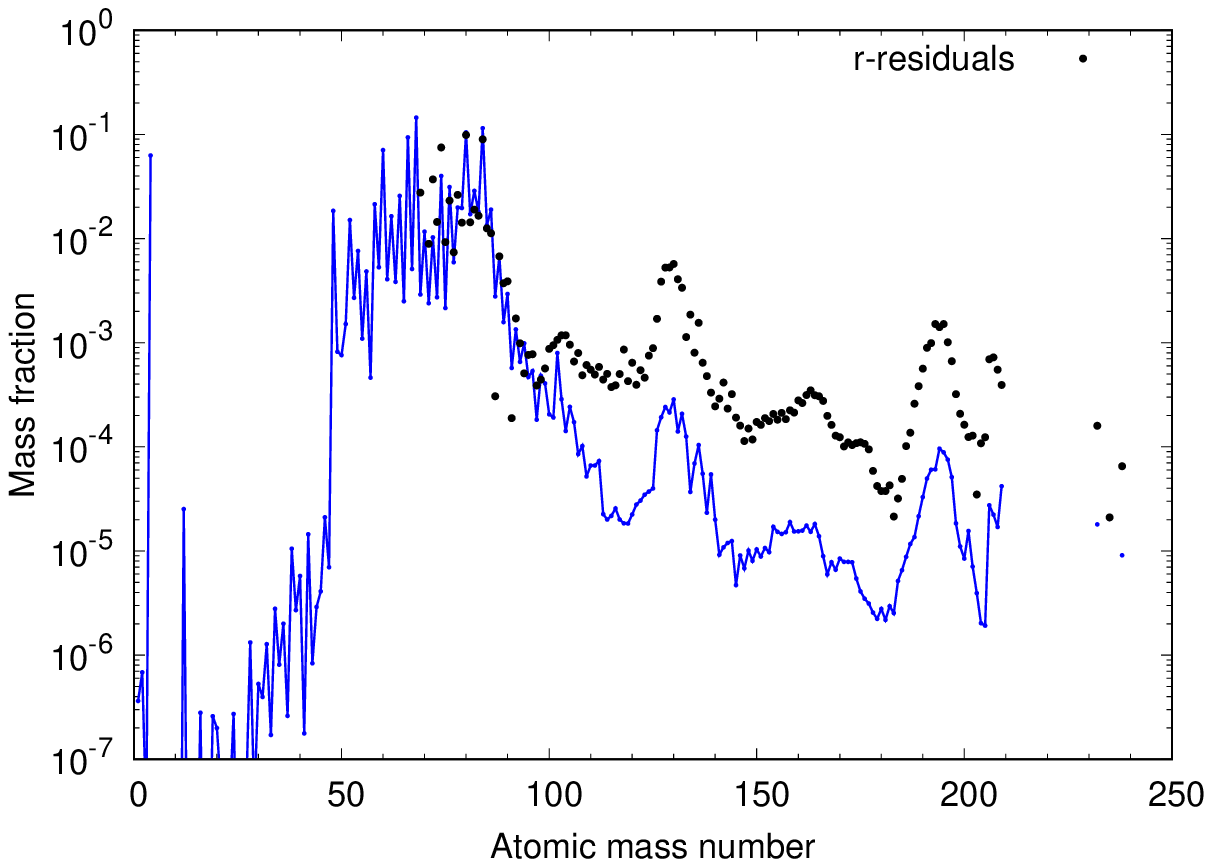} &
    \includegraphics[width=0.48\linewidth]{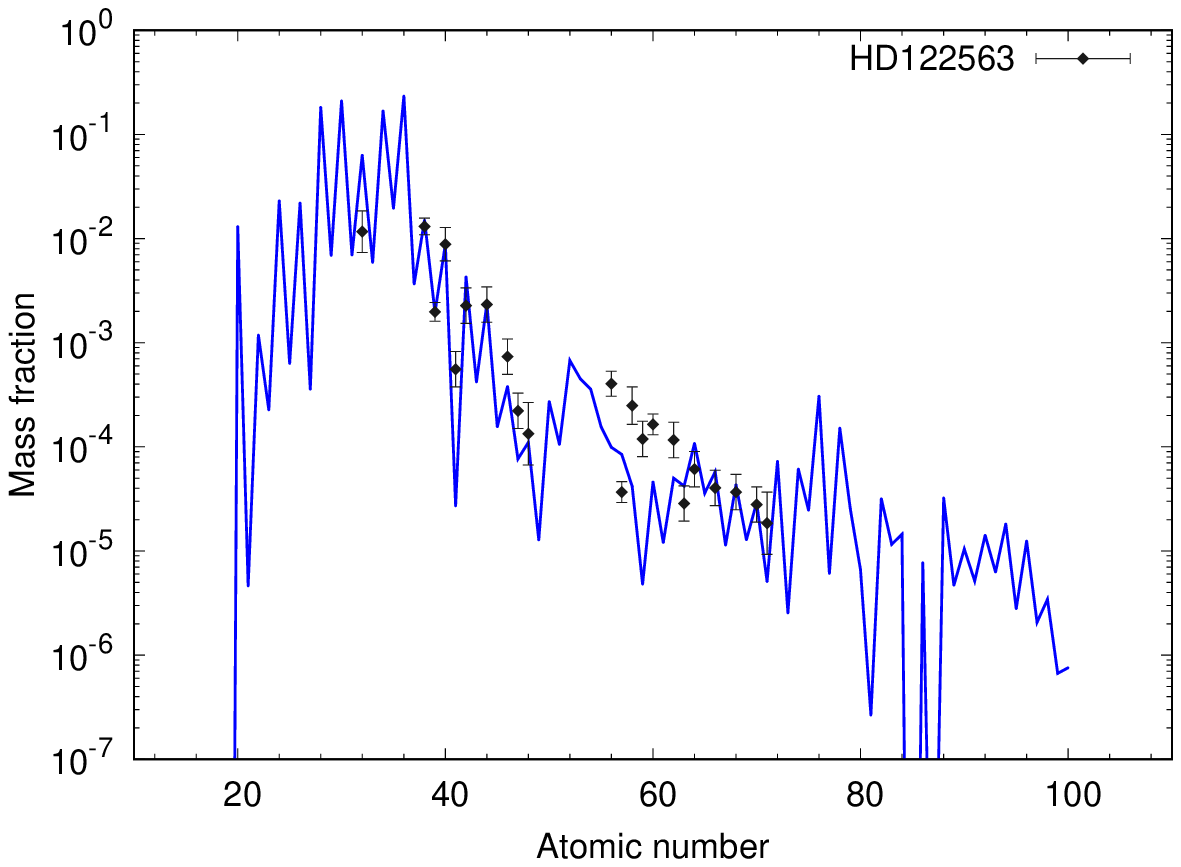}  \\
    \end{tabular}
\caption{
  \label{fig:abun}
  Left: final abundances of our L model as a function of mass number.
  Black circles show the $r$-process residual pattern \citep{Prantzos2020}, which are scaled to match those for the L model at $A=88$.
  Right: abundances at $t=1.5$ days as a function of atomic number.
  Abundances of an $r$-process-deficient star HD 122563 (diamonds, \citealp{Honda2006}; Ge from \citealp{Cowan2005}; Cd and Lu from \citealp{Roederer2012}) are also shown for comparison, which are scaled to match those for the L model at $Z=40$.
}
\end{center}
\end{figure*}
\begin{deluxetable*}{ccccccccc}[ht]
\tablewidth{0pt}
\tablecaption{Mass fractions of selected elements in the L model.
 The top and bottom rows show the final abundances and those at $t=1.5$ days, respectively.}
\label{tab:abun}
\tablehead{
  $X$(Ca) & $X$(Sr) & $X$(Y) & $X$(Zr) & $X$(Ba) & $X$(La) & $X$(Ce) & $X$(Th) & $X$(La+Ac)$^a$ 
}
\startdata
$1.8\times 10^{-2}$ & $6.8\times 10^{-3}$ & $1.6\times 10^{-3}$ & $6.4\times 10^{-2}$ & $1.5\times 10^{-4}$ & $5.4\times 10^{-5}$ & $3.1\times 10^{-5}$ & $1.8\times 10^{-5}$ & $4.9\times 10^{-4}$ \\ 
$1.3\times 10^{-2}$ & $1.5\times 10^{-2}$ & $2.0\times 10^{-3}$ & $8.8\times 10^{-3}$ & $9.9\times 10^{-5}$ & $8.5\times 10^{-5}$ & $4.2\times 10^{-5}$ & $1.0\times 10^{-5}$ & $6.7\times 10^{-4}$ \\ \hline 
\enddata
\tablecomments{
$^a$ Sum of mass fractions for lanthanides ($Z=57$--71) and actinides ($Z=89$--100).
}
\end{deluxetable*}

To evaluate the strength of bound-bound transitions in NS merger ejecta, we essentially need atomic data.
In this paper, a dataset of transition wavelength, energy level of transition, and transition probability is referred to a line list.

For theoretical calculations of kilonova light curves, atomic data obtained from theoretical calculations have been often used \citep[e.g.,][]{Kasen2013, Tanaka2018, Tanaka2020, Fontes2020, Banerjee2020}.
This is useful in terms of completeness of the transition lines, because the opacity of ejecta should be correctly evaluated for light curve calculations. 
However, while such theoretical data give a reasonable estimate for the total opacity, they are not necessarily accurate in transition wavelengths, and thus, not suitable for element identification.

\citet{Domoto2021} used a latest line list constructed from the Vienna Atomic Line Database \citep[VALD;][]{Piskunov1995, Kupka1999, Ryabchikova2015} to focus on the imprint of elemental abundances in kilonova spectra.
This database is suitable for identifying lines because the atomic data are calibrated with experiments and semi-empirical calculations.
Since most spectroscopic experiments have been conducted in the optical range, there is enough data to investigate spectral features at optical wavelengths.
However, such experimental line list is not necessarily complete in the NIR region.

Here, we propose a new scheme to take the advantages of both line lists.
By using a {\it complete} line list constructed from theoretical calculations, we first identify which elements can show strong transitions under the physical conditions of NS merger ejecta.
Then, we calibrate the theoretical energy levels with experimental data and construct an {\it accurate} line list for the selected ions with strong transitions.
In this way, we construct a {\it hybrid} line list that is complete for weak transitions and accurate for strong transitions, which are important for element identification.

\subsection{Candidate species}
\label{sec:candidate}
To investigate which elements can become absorption sources in kilonova photospheric spectra, we systematically calculate the strength of bound-bound transitions for given density, temperature, and element abundances.
The strength of a line is approximated by the Sobolev optical depth \citep{Sobolev1960} for each bound-bound transition,
\begin{eqnarray}
	\tau_l &=& \frac{\pi e^2}{m_e c} n_{i, j, k} t \lambda_l f_l  \nonumber \\
		&=& \frac{\pi e^2}{m_e c} n_{i, j} t \lambda_l f_l \frac{g_k}{g_0} e^{-\frac{E_k}{kT}},
	\label{eq:tau}
\end{eqnarray}
in homologously expanding ejecta. 
The Sobolev approximation is valid for the matter with a high expansion velocity and a large radial velocity gradient.
Here, $n_{i, j, k}$ is the number density of ions at the lower level of a transition ($i$-th element, $j$-th ionization stage, and $k$-th excited state), $f_l$ and $\lambda_l$ are the oscillator strength and the transition wavelength, $g_0$ is the statistical weight at the ground state, and $g_k$ and $E_k$ are the statistical weight and the lower energy level of a bound-bound transition, respectively.
As in previous work on kilonovae \citep[e.g.,][]{BK2013, TH2013}, we assume local thermodynamic equilibrium (LTE); 
we solve the Saha equation to obtain ionization states, and assume Boltzmann distribution for the population of excited levels, which appears in Equation \ref{eq:tau} (see \citealp{Pognan2022a} for non-LTE effects).

For the abundances in the ejected matter from a NS merger, we use the same model as in \citet{Domoto2021} based on a multi-component free-expansion (mFE) model of \citet{Wanajo2018}.
Here, we use the Light (L) model as our fiducial model (the left panel of Figure \ref{fig:abun}), which exhibits a similar abundance pattern to that of metal-poor stars with weak $r$-process signature  \citep[e.g., HD 122563,][the rihgt panel of Figure \ref{fig:abun}]{Honda2006}.
This is motivated by the fact that the blue emission of GW170817/AT2017gfo at a few days after the merger is suggested to have stemmed from the ejecta component dominated by relatively light $r$-process elements \citep[e.g.,][]{Arcavi2017, Nicholl2017, Tanaka2017, Tanaka2018}.
Although the model includes the abundances with the atomic number of $Z=1$--110, we use only the abundances at $t=1.5$ days with $Z=20$--100 in our calculations as shown in the right panel of Figure \ref{fig:abun}.
The mass fractions of elements relevant to this study is summarized in Table \ref{tab:abun}.
Note that the calculated abundances, which are re-computed with an updated nucleosynthesis code \citep{Fujibayashi2020b, Fujibayashi2022}, slightly differ (within a factor of 2 at 1.5 days) from those presented in \citet{Domoto2021}.
We have confirmed that our new nucleosynthetic abundances give almost the same results with those in \citet{Domoto2021}.

For the atomic data, we use a theoretical line list from \citet{Tanaka2020}.
This line list was constructed for the elements with $Z =30$--88 from neutral atoms up to triply ionized ions by systematic atomic structure calculations using the HULLAC code \citep{HULLAC}.
Since transition data of this line list are not necessarily accurate in terms of wavelengths, we here only study the species of ions to produce strong transitions.

Note that the theoretical line list does not include the data for actinides ($Z=89$--100) due to the difficulty of atomic structure calculations \citep{Tanaka2020}.
While actinides just work as zero-opacity sources without atomic data, we keep these elements in the list to discuss the spectral features of actinides in Section \ref{sec:actinide}.

\begin{figure*}[th]
  \begin{center}
    \begin{tabular}{cc}
    \includegraphics[width=0.48\linewidth]{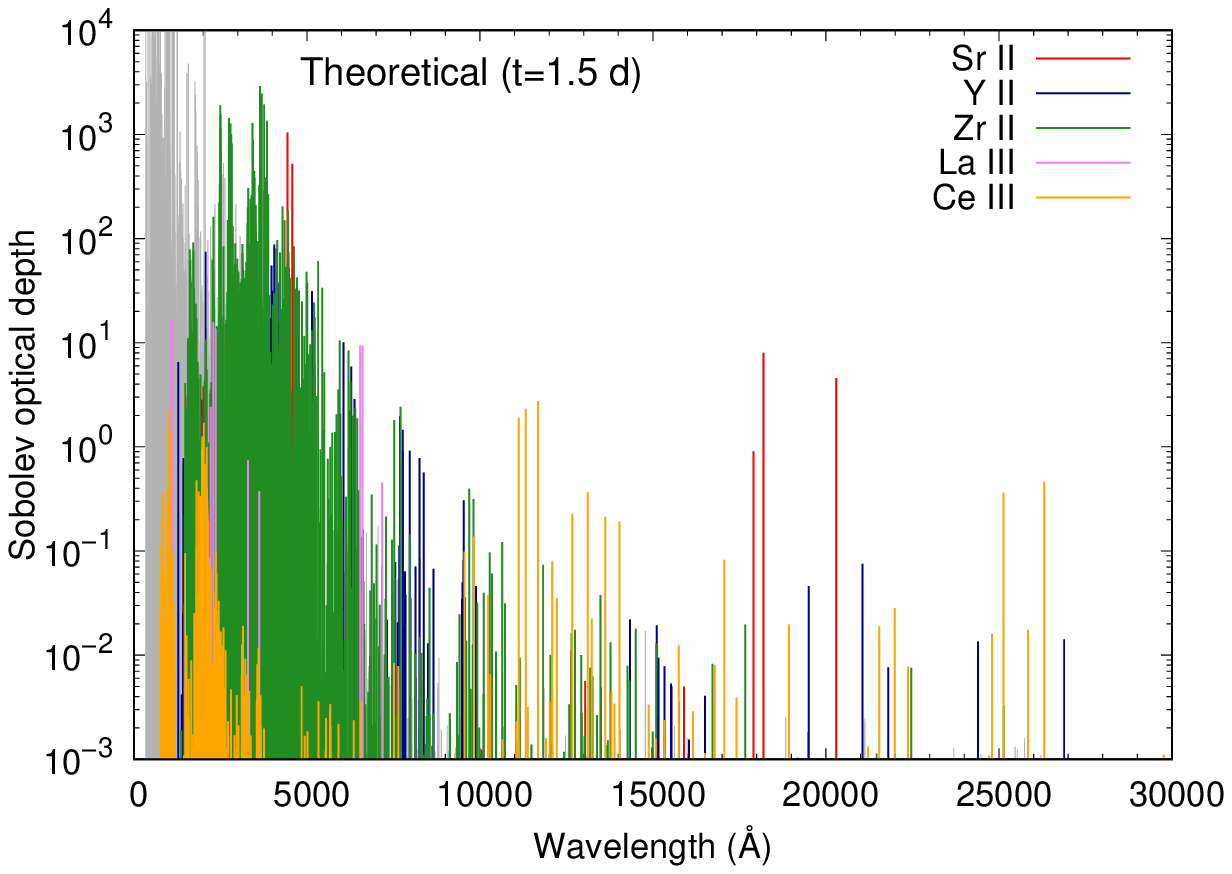} &
    \includegraphics[width=0.48\linewidth]{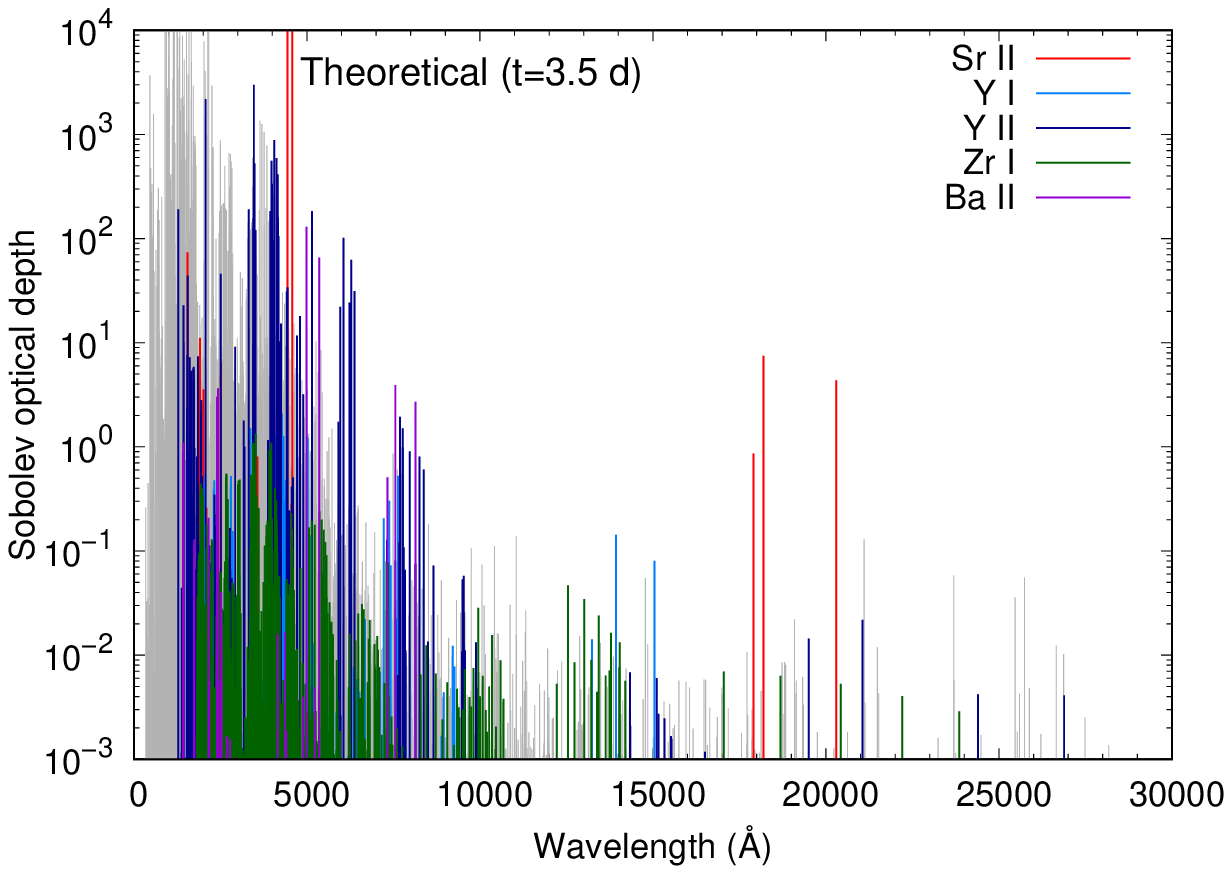}  \\
    \includegraphics[width=0.48\linewidth]{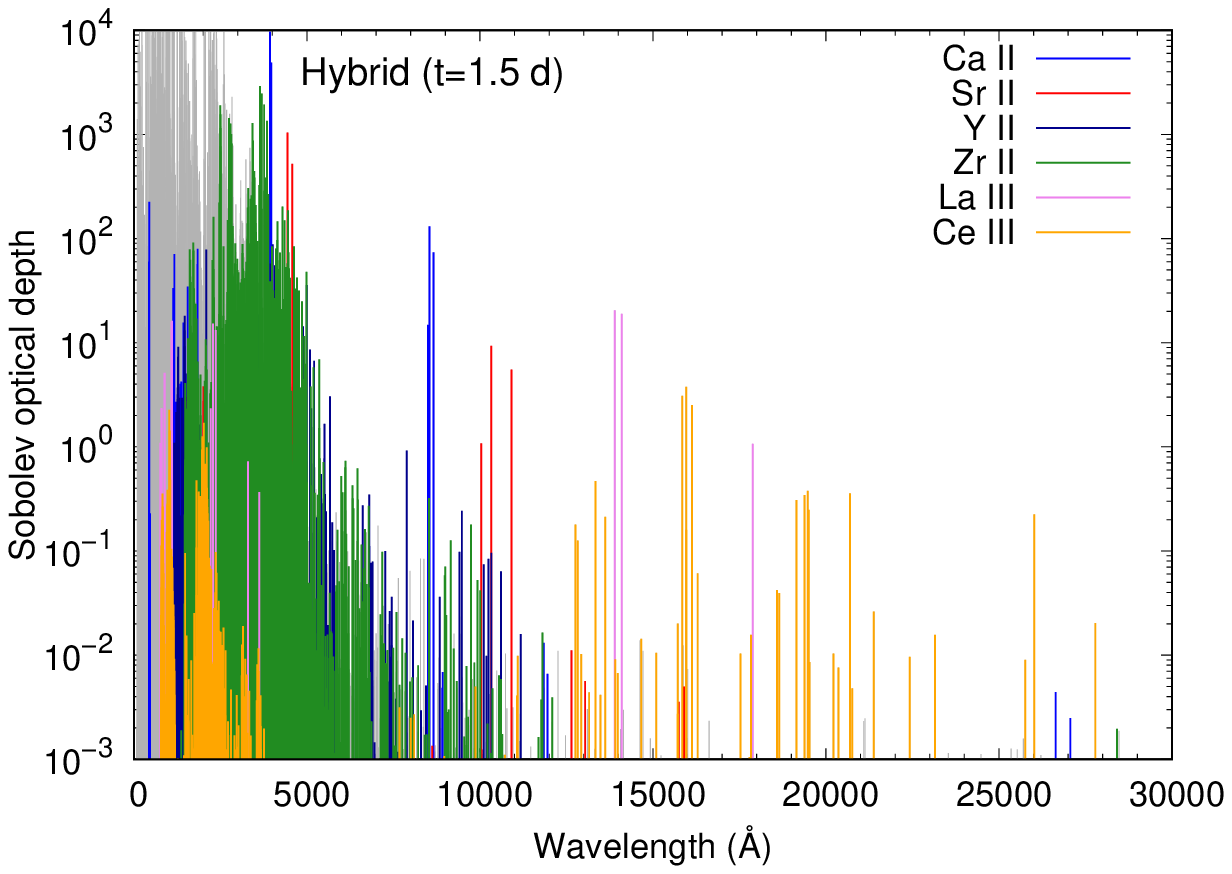} &
    \includegraphics[width=0.48\linewidth]{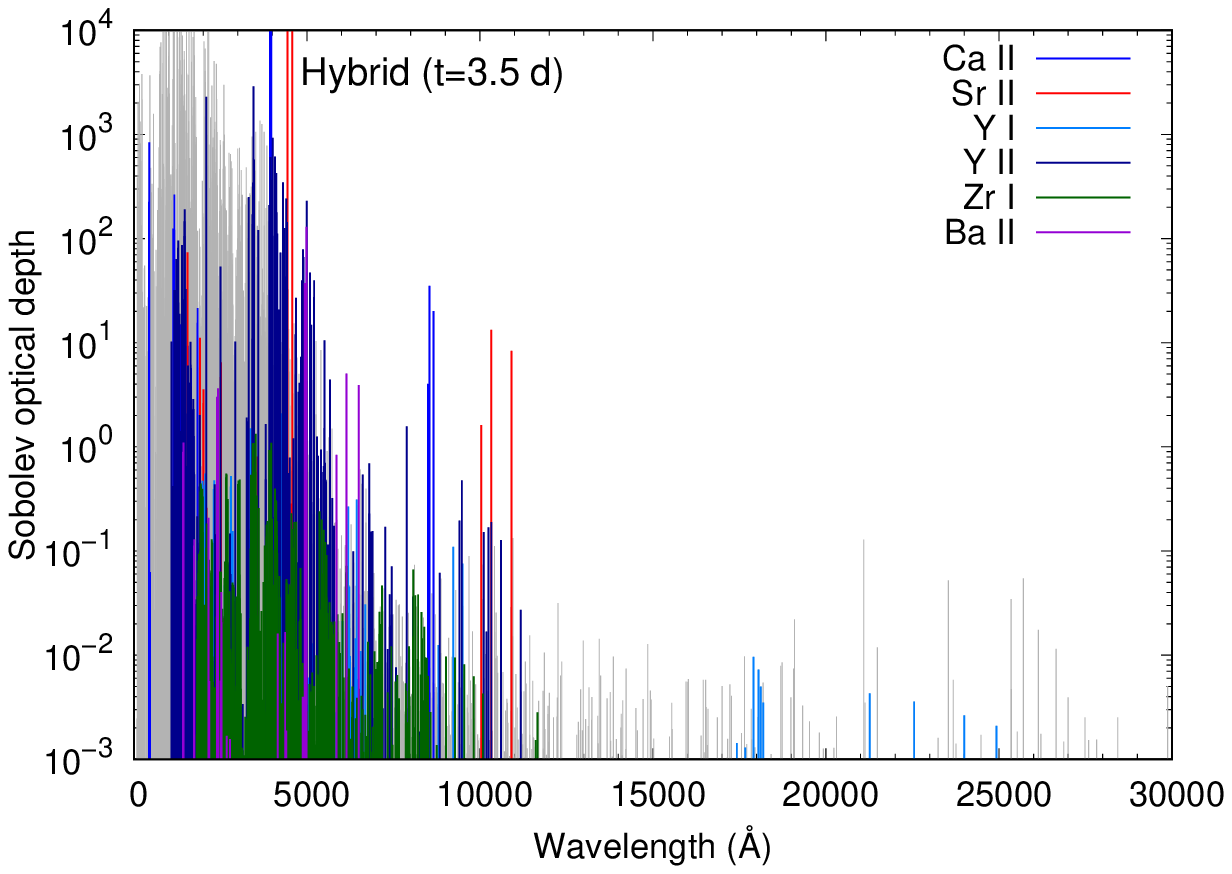}  \\
    \end{tabular}
\caption{
  \label{fig:tau}
  Sobolev optical depth of bound-bound transitions for the L model calculated with the theoretical line list \citep[$Z = 30$--88, top]{Tanaka2020}
  and those calculated with the hybrid line list ($Z = 20$--88, bottom, Section \ref{sec:hybrid}).
  The ions with large contributions are shown with colors.
  The left panels show the results with the density of $\rho=10^{-14}$ g~cm$^{-3}$ and the temperature of $T=5000$ K at $t=1.5$ days, 
  while the right panels show those with $\rho=10^{-15}$ g~cm$^{-3}$ and $T=3000$ K at $t=3.5$ days.
}
\end{center}
\end{figure*}

The strength of bound-bound transitions at $t=1.5$ and 3.5 days for the L model is shown in the top panels of Figure \ref{fig:tau}.
We evaluate the Sobolev optical depth for the density of $\rho=10^{-14}$ g~cm$^{-3}$ and the temperature of $T=5000$ K at $t=1.5$ days, and $\rho=10^{-15}$ g~cm$^{-3}$ and $T=3000$ K at $t=3.5$ days.
These are typical values in the line forming region when we adopt the abundance distribution of the L model (see Figure \ref{fig:temp}).

We find that, among all the elements, Y II, Zr II, La III, and Ce III lines are as strong as the Sr II triplet lines under the condition at $t=1.5$ days.
On the other hand, under the condition at $t=3.5$ days, Y I, Zr I, and Ba II lines appear instead of Zr II, La III, and Ce III lines.
Although the individual lines of Zr I are not by far the strongest, they can be important absorption sources due to the fact that the multiple lines show comparable strength at a certain wavelength range.

The behavior of the strength of the lines can be understood as the dependence of the Sobolev optical depths on temperature and density \citep[Figure 4 of][]{Domoto2021}.
When the density is $\rho=10^{-14}$--$10^{-15}$ g~cm$^{-3}$, the strength of Sr II triplet lines does not show a large difference between $T=3000$ and 5000 K.
This is determined by the combination of the ionization fraction and the population of the levels at these temperatures.
While the lines of singly ionized Sr ($Z = 38$) are not largely affected in this temperature range, the lines of neutral Y ($Z = 39$) and Zr ($Z = 40$) appear at $T = 3000$ K.
This is due to the slightly higher ionization potentials of Y and Zr.
On the other hand, the strength of Ce III lines largely changes in this temperature range reflecting the ionization fraction.
Most of Ce atoms are singly ionized at $T=3000$ K, and thus, the lines of Ce III become weaker at lower temperature.
The behavior of Ba II ($Z = 56$) and La III ($Z = 57$) lines can be explained in a similar way to that of Ce III ($Z = 58$).

\subsection{Atomic properties}
\label{sec:atomic}
\begin{figure*}[ht]
  \begin{center}
    \begin{tabular}{cc}
    \includegraphics[width=0.48\linewidth]{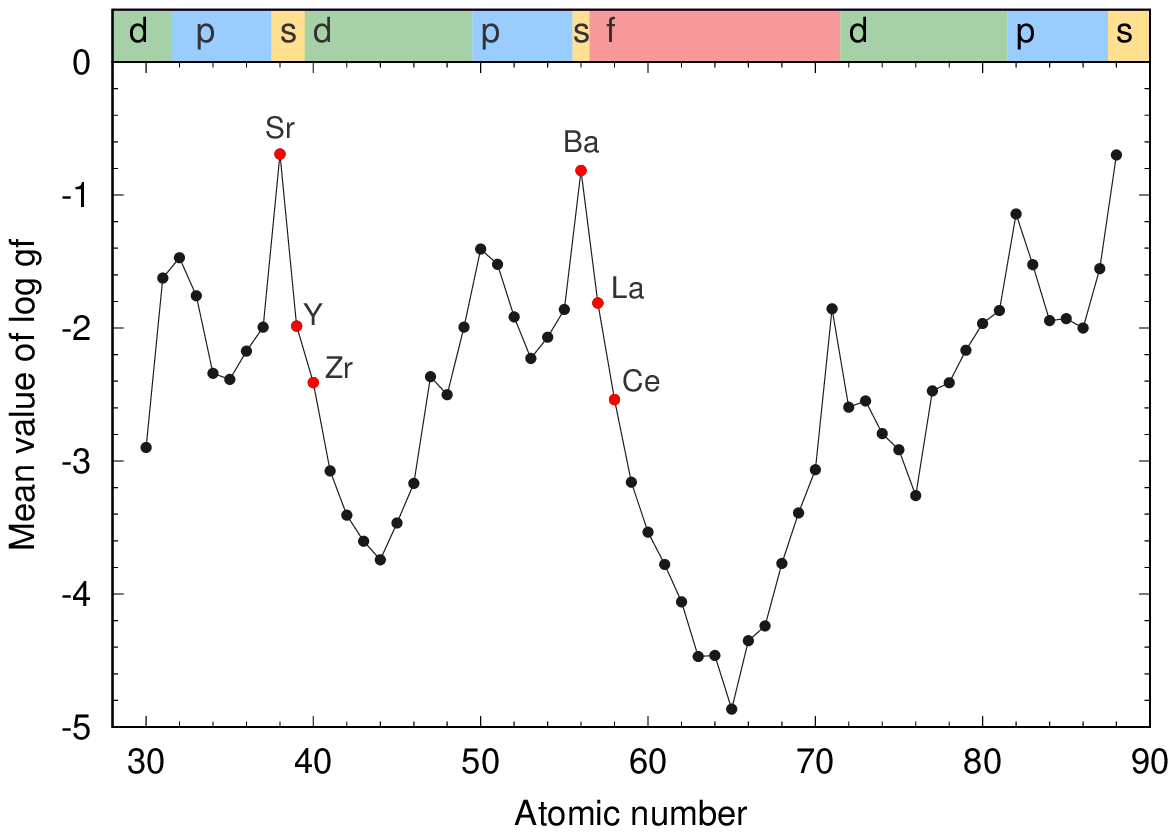} &
    \includegraphics[width=0.48\linewidth]{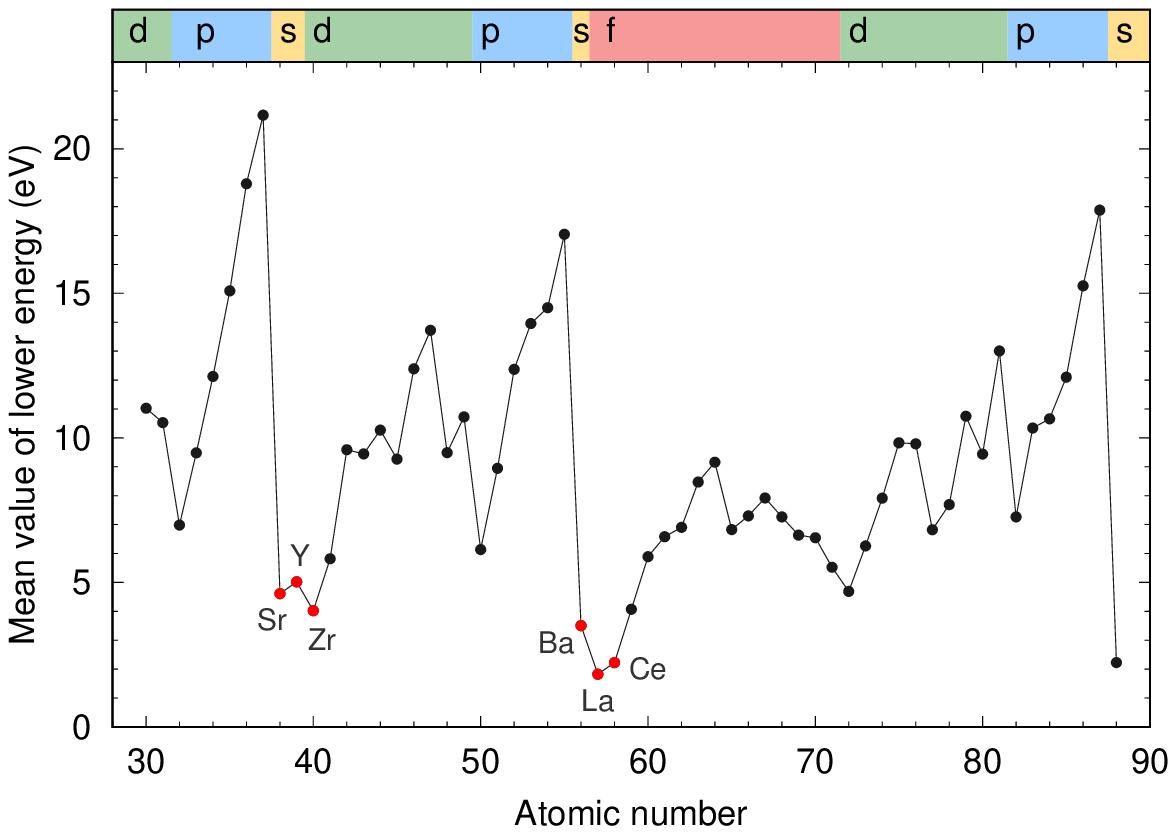}  \\
    \end{tabular}
\caption{
  \label{fig:vsZ}
  Mean values of log $gf$ (left) and of lower energies of transitions (right) for all the lines of singly ionized ions as a function of atomic number.
  Colors at the top show the valence shells with the orbital angular momenta $l$ for singly ionized ions.
  Red circles indicate the elements that produce strong transitions in our analysis (see the text).
}
\end{center}
\end{figure*}
\begin{figure}[ht]
  \begin{center}
    \includegraphics[width=\linewidth]{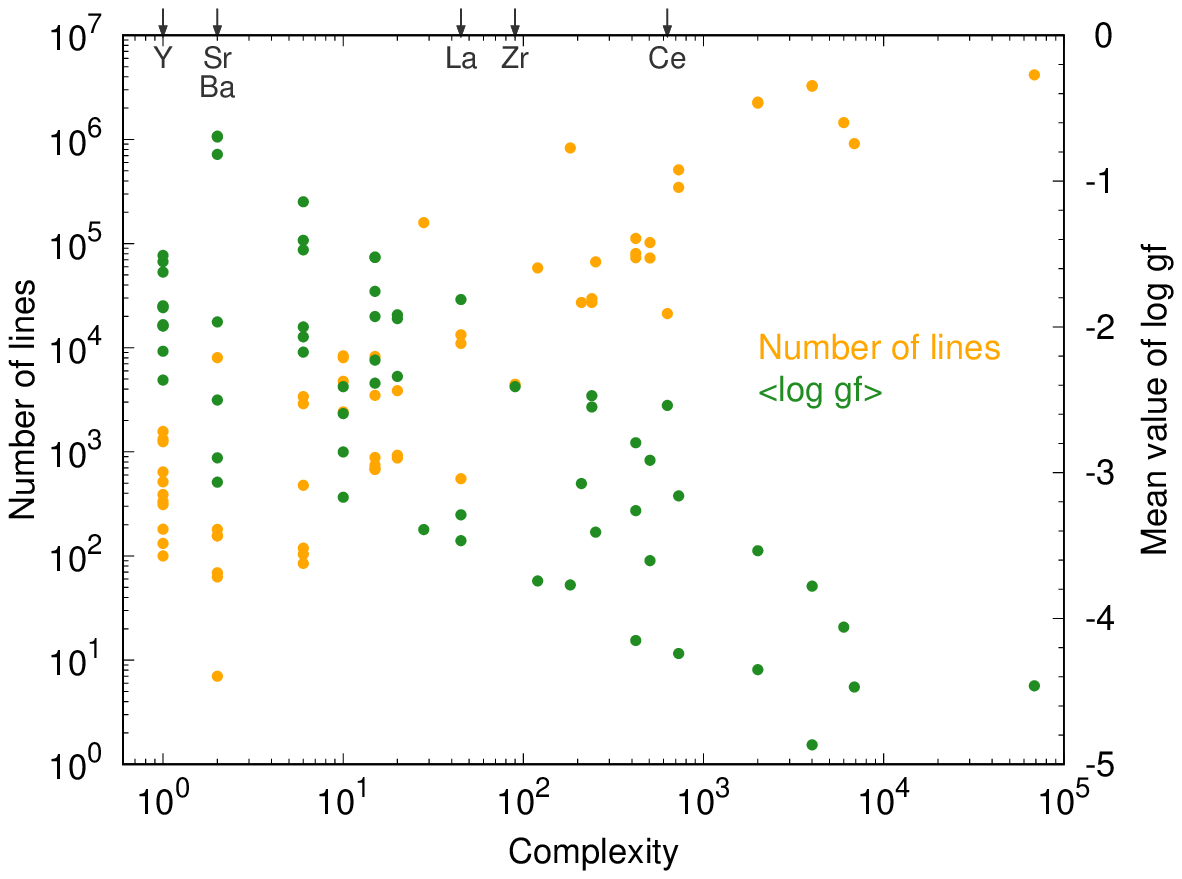}
\caption{
  \label{fig:vsC}
  Number of lines (orange) and mean value of log $gf$ (green) for singly ionized ions as a function of complexity.
}
\end{center}
\end{figure}

Although our line list includes all elements with $Z=30$--88, our results indicate that only a few elements, such as Sr, Y, Zr, Ba, La, and Ce can become strong absorption sources in the spectra.
The reason can be interpreted by the atomic properties of these elements.
From Equation \ref{eq:tau}, necessary conditions for a given lines to become strong are that
(1) the transition probability ($g_k f_l$) is high and 
(2) the lower energy level of the transition ($E_k$) is low (\ie level population is high).

The left panel of Figure \ref{fig:vsZ} shows the mean values of log $gf$ for all the lines of singly ionized ions as a function of atomic number.
The mean $gf$-values show the pattern according to the orbital angular momentum $l$ of the valence shell.
This is more understandable when we use the complexity of a given ion, defined as \citep{Kasen2013}:
\begin{equation}
	C = \Pi_m \frac{g_m!}{n_m! (g_m-n_m)!},
	\label{eq:comp}
\end{equation}
where $g=2(2l+1)$ is the number of magnetic sublevels in the subshell with orbital angular momentum $l$, and $n_m$ is the number of electrons in the $nl$-orbital labeled $m$.
The complexity indicates how dense the energy levels are packed, and takes the maximal value when filling half the closed-shell \citep[Figure 1 of][]{Kasen2013}.
Figure \ref{fig:vsC} shows the number of lines and the mean value of log $gf$ for singly ionized ions as a function of complexity.
We find that the complexity shows a positive correlation with the number of lines.
It is natural because the number of transition combinations increases for larger complexity, \ie denser energy levels.
We also find that the complexity shows a negative correlation with the mean $gf$-value.
This can be understood by the sum rule of oscillator strength;
when the ion has a larger number of transitions, the oscillator strength of each line tends to be smaller.
Other ionization states are not presented in Figure \ref{fig:vsC}, but show similar trends.
Therefore, the ions with relatively low complexity are likely to have transitions with relatively high $gf$-values.

The right panel of Figure \ref{fig:vsZ} shows the mean values of the lower energy levels of transitions for all the lines of singly ionized ions as a function of atomic number.
The mean lower energy level of transitions tends to be pushed up toward higher energy as atomic number increases in an electron shell series.
Other ionization states also show similar behavior.
As spin-orbit interaction energy strongly depends on atomic number, energy-level spacing at a certain shell increases with atomic number.
Also, electron orbital radii become smaller as atomic number increases, so that electron-electron interaction energies become higher for larger atomic number.
As a result, distribution of energy levels becomes wider for larger atomic number at a given shell \citep{Tanaka2020}.
Therefore, the ions with smaller atomic numbers in a certain period on the periodic table tend to have transitions from low-lying energy levels.

Consequently, it is natural that Sr, Y, Zr, Ba, La, and Ce show strong lines.
According to the two properties mentioned above, the ions on the left side of the periodic table are anticipated to show strong lines.
In fact, all of the elements showing strong lines belong to group 2 to 4 on the periodic table;
in other words, they have a relatively small number of valence electrons (low complexity) and relatively low-lying energy levels.

Among these ions, La III and Ce III show strong lines at the NIR wavelengths (Figure \ref{fig:tau}).
This can be understood from the properties of lanthanide elements.
While lanthanide elements are characterized by having the 4$f$-electrons, the configurations of low-lying energy levels for lanthanides also involve the outer 5$d$ and 6$s$ shells.
This means that the energy scales of 4$f$, 5$d$, and 6$s$ orbitals for lanthanides are similar;
in other words, the energy differences between these orbitals are small.
In fact, the strong transitions of La III and Ce III at the NIR wavelengths involve an electron jump between 4$f$ and 5$d$ orbitals (see Table \ref{tab:LaIII} and \ref{tab:CeIII}).
Thus, it is natural that La III and Ce III lines tend to appear in the NIR region.

\subsection{Construction of hybrid line list}
\label{sec:hybrid}
\begin{deluxetable*}{lcccc}[ht]
\tablewidth{0pt}
\tablecaption{Summary of the hybrid line list.}
\label{tab:list}
\tablehead{
  & Element$^{a}$ & Ion & \multicolumn{2}{c}{Reference} \\
  &                         &       &  Levels  &  Transitions
}
\startdata
Baseline 	 & $Z=20$--29   & I--IV & \multicolumn{2}{c}{1} \\
		 & $Z=30$--88   & I--IV & \multicolumn{2}{c}{2} \\
		 & $Z=89$--100 &         & \multicolumn{2}{c}{no lines} \\ \hline
Calibrated ions & Sr ($Z=38$) &   II  &  3  & 1 \\
			& Y ($Z=39$)  & I, II &  3  & 1 \\
			& Zr ($Z=40$) & I, II &  3  & 4 \\
			& Ba ($Z=56$) &  II  &  3  & 1 \\
			& La ($Z=57$) &  III &  3  & 2 \\
			& Ce ($Z=58$) &  III &  3  & 2 \\ \hline
Section \ref{sec:actinide}       & Th ($Z=90$) & III &  \multicolumn{2}{c}{1 (optical)} \\
					     &                     &      &  3  &  3, 5 (NIR)$^b$ \\
\enddata
\tablecomments{
$^a$ Elements of $Z=20$--100 are used for all the calculations in this paper.\\
$^b$ Relative intensities are used to estimate the transition probabilities (see Section \ref{sec:actinide}). \\
References: (1) VALD \citep{Piskunov1995, Kupka1999, Ryabchikova2015}; (2) \citet{Tanaka2020}; (3) NIST Atomic Spectral Database \citep{NIST_ASD}; (4) Kurucz's atomic data \citep{Kurucz2018}; (5) \citet{Engleman2003}.
}
\end{deluxetable*}
We find that Y, Zr, Ba, La, and Ce, as well as Sr, can show strong lines under the physical conditions of NS merger ejecta.
To enable us to inspect absorption lines produced by these elements in kilonova photospheric spectra, we calibrate the energy levels and resulting transition wavelengths of theoretical atomic data with experimental data.
The details for calibration procedures are given in Appendix \ref{sec:lsj}.

To use the calibrated lines in radiative transfer simulations (Section \ref{sec:spectra}), we need their transition probabilities.
After calibrating the energy levels and wavelengths, the transition probabilities of the calibrated lines are taken from the VALD database \citep{Piskunov1995, Kupka1999, Ryabchikova2015} or Kurucz's atomic data \citep{Kurucz2018} if the lines are listed, otherwise theoretical values from \citet{Tanaka2020} are adopted.
The general validity of the theoretical transition probabilities is discussed in Appendix \ref{sec:lsj}.
The transitions with the theoretical transition probabilities are summarized in Table \ref{tab:LaIII}--\ref{tab:CeIII} (Appendix \ref{sec:lsj}).

Finally, we construct a hybrid line list by combining the VALD database for $Z = 20$--29 and the results of atomic calculations from \citet{Tanaka2020} for $Z = 30$--88.
Among the data of $Z = 30$--88, strong transitions of Sr II, Y I, Y II, Zr I, Zr II, Ba II, La III, and Ce III are replaced with those calibrated with experimental data.
The information for the hybrid line list is summarized in Table \ref{tab:list}.
The bottom panels of Figure \ref{fig:tau} show the strength of bound-bound transitions at $t=1.5$ and 3.5 days calculated with the new hybrid line list.
In these panels, the wavelengths of strong transitions are accurate as they are calibrated with experimental data.
We find that La III and Ce III lines become strong absorption sources at the NIR wavelengths at $t=1.5$ days.
For Y I, Y II, Zr I, Zr II, and Ba II, most of the wavelengths of relatively strong lines are found to be placed in the the optical region ($\lambda < 10000$ {\AA}).
Thus, their lines are unlikely to produce absorption features in the NIR region, although they may be important absorption sources at the optical wavelengths (see Section \ref{sec:spectra}).

\begin{figure}[ht]
  \begin{center}
     \includegraphics[width=0.95\linewidth]{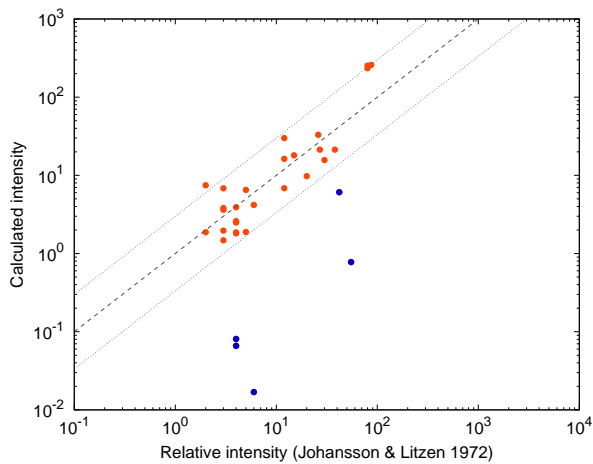}
\caption{
  \label{fig:intensity}
  Comparison of intensities for NIR Ce III lines between those calculated with theoretical $gf$-values and those measured by experiments \citep{Johansson1972}.
  Gray dashed and dotted lines correspond to perfect agreement and deviations by a factor of 3, respectively.
  Blue circles indicate the lines whose theoretical $gf$-values are underestimated more than a factor of 3.
}
\end{center}
\end{figure}

We adopt the theoretical transition probabilities when they are not available in the VALD database or the Kurucz's line list.
To validate the accuracy of theoretical $gf$-values for NIR lines, we use the relative intensities of Ce III lines.
\citet{Johansson1972} measured the emission lines of Ce III at the NIR wavelengths in a laboratory and showed the relative intensities of measured lines.
Assuming LTE for ionization and excitation, the intensity of an emission line can be calculated as 
\begin{eqnarray}
	I &=& b\ g_u A\ e^{-\frac{E_u}{kT}} \nonumber \\
	  &=& b \frac{8\pi e^2}{m_e c\lambda_l^2}g_l f_l e^{-\frac{E_u}{kT}},
	\label{eq:I}
\end{eqnarray}
where $A$, $g_u$, and $E_u$ are the Einstein's A coefficient, the statistical weight, and the energy level of the upper level for a transition, respectively, and $b$ is a constant depending on the ion species.
Since plasma in experiments is typically in LTE due to the high density of ions \citep{Kielkopf1971}, we can use this formula to evaluate the $gf$-values of Ce III lines.

Comparison between the intensities of Ce III lines calculated with the theoretical transition probabilities and those measured by experiments \citep{Johansson1972} is shown in Figure \ref{fig:intensity}.
We adopt the temperature of $T=12000$ K, which is typical plasma temperature in experiments \citep{Kielkopf1971}.
Here a normalization factor $b$ is set so that the calculated values are close to the experimental values.
We find that the calculated and experimentally measured intensities are  in good agreement except for a few lines.
Since in the situation we consider ($E_u \lesssim 2$ eV) the intensities are mainly determined by transition probabilities, the trend suggests that our theoretical $gf$-values of Ce III lines are reasonable.
We note that, although the $gf$-values of a few lines should be higher than our estimates (blue circles in Figure \ref{fig:intensity}), they are relatively weak and do not affect our conclusions.
Nevertheless, to determine the exact values of transition probabilities for these lines, more experimental and observational calibrations are necessary in the NIR region.

\section{Synthetic spectra}
\label{sec:spectra}
\subsection{Methods}
In this section, we calculate realistic synthetic spectra of kilonovae by using the new hybrid line list.
We use a wavelength-dependent radiative transfer simulation code \citep{TH2013, Tanaka2014, Tanaka2017, Tanaka2018, Kawaguchi2018, Kawaguchi2020}.
The photon transfer is calculated by the Monte Carlo method. 
To compute the opacity for bound-bound transitions, we adopt the expansion opacity \citep{Karp1977} and use the formula from \citet{EastmanPinto1993}:
\begin{equation}
	\kappa_{\exp}(\lambda)=\frac{1}{c t \rho} \sum_l \frac{\lambda_l}{\Delta \lambda}(1-e^{-\tau_l}) ,
	\label{eq:kappa}
\end{equation}
where $\tau_l$ is the Sobolev optical depth (Equation (\ref{eq:tau})). 
In the equation, the summation is taken over all transitions within a wavelength bin $\Delta \lambda$ (see below).
The Sobolev optical depth is evaluated by assuming LTE for ionization and excitation as in Section \ref{sec:candidate}.

For the atomic data, we use the new hybrid line list constructed in Section \ref{sec:hybrid}.
The hybrid line list still includes weak transitions whose wavelengths are not necessarily accurate.
To avoid the substantial effects of these lines to spectra, we adopt a wide wavelength grid for the opacity calculation with the atomic data from theoretical calculations (\ie lines for $Z=30$--88).
The wavelength grid is typically set to $\Delta \lambda=10$ {\AA} \citep{TH2013}, but here a 20 times wider grid is adopted for the theoretical line list.
This smears out the individual effect of each line on the bound-bound opacity.
We also performed the same opacity calculations with the typical fine wavelength grid, and confirmed that the resultant total opacity is almost unchanged.
For the accurate transitions (\ie lines for $Z=20$--29 and calibrated lines), we adopt $\Delta \lambda=10$ {\AA}.
By combining the opacity calculated with the theoretical atomic data and the strong transitions calculated with the accurate data, we are able to discuss whole spectral features, \ie an overall shape, absorption lines, and their time evolution.

\begin{figure}[th]
  \begin{center}
    \includegraphics[width=\linewidth]{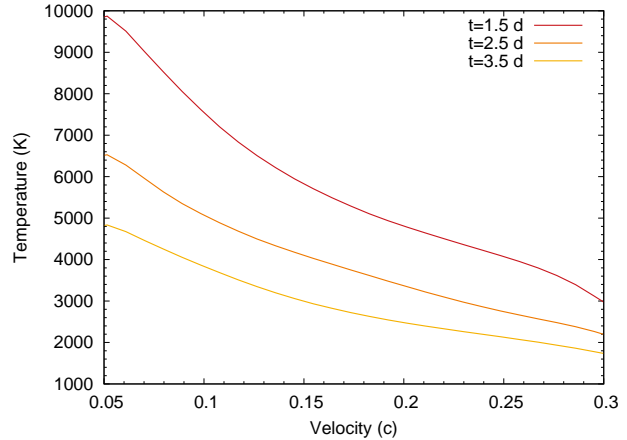}
\caption{
  \label{fig:temp}
  Temperature structure of the ejecta at $t=1.5$, 2.5, and 3.5 days after the merger.
}
\end{center}
\end{figure}

In the radiative transfer code, the temperature in each cell is determined by the photon flux \citep{Lucy2003, TH2013}.
The photon intensity is evaluated as
\begin{equation}
	J_\nu d\nu = \frac{1}{4\pi \Delta t V} \sum_{d\nu}\epsilon ds,
\end{equation}
where $\epsilon$ is the comoving-frame energy of a photon packet.
The temperature is estimated by assuming that the wavelength-integrated intensity $\langle J \rangle = \int J_\nu d \nu$ follows the Stefan-Boltzmann law, \ie
\begin{equation}
	\langle J \rangle = \frac{\sigma}{\pi}T_R^4.
\end{equation}
The kinetic temperature of electrons $T_e$ is assumed to be the same as the radiation temperature $T_R$, \ie $T=T_e=T_R$ under LTE.

For the ejecta density structure, we assume a single power law ($\rho \propto r^{-3}$) for the velocity range of ejecta $v=0.05$--$0.3\ c$ \citep[e.g,][]{Metzger2010}.
The total ejecta mass is set to be $\Mej=0.03\Msun$, which is suggested to explain the observed luminosity of AT2017gfo \citep[e.g.,][]{Tanaka2017, Kawaguchi2018}.
For the abundance distribution, we use the same model (L model) as described in Section \ref{sec:candidate}.
The heating rate of radioactive nuclei as a function of time is consistently computed for this model.
The thermalization efficiency of $\gamma$-rays and radioactive particles follows the analytic formula given by \citet{Barnes2016}.
The resulting temperature structure of ejecta is shown in Figure \ref{fig:temp}.

\subsection{Results}
\label{sec:spectra result}
\begin{figure}[th]
  \begin{center}
    \includegraphics[width=1.02\linewidth]{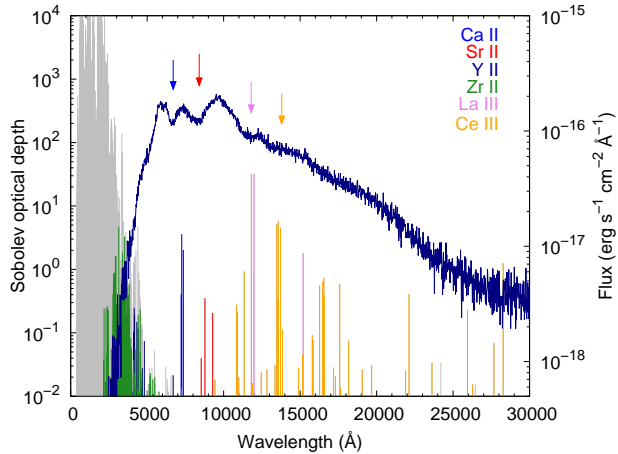}
\caption{
  \label{fig:flux}
  Synthetic spectrum (blue curve) and Sobolev optical depth of each transition (vertical lines) at $t=1.5$ days.
  We plot the Sobolev optical depths of spectroscopically accurate lines in the ejecta at $v = 0.16\ c$.
  The positions of lines are blueshifted according to $v=0.16\ c$.
  The temperature in the ejecta at $v= 0.16\ c$ is $T\sim6000$ K.
}
\end{center}
\end{figure}

Figure \ref{fig:flux} shows the synthetic spectrum at $t=1.5$ days after the merger.
To show the contribution of different elements, we also plot the Sobolev optical depths in the ejecta at $v=0.16\ c$.
The wavelengths of lines are blueshifted according to $v=0.16\ c$.
Note that we plot only spectroscopically accurate lines that contribute to absorption features.

\begin{figure*}[ht]
  \begin{center}
    \includegraphics[width=0.9\linewidth]{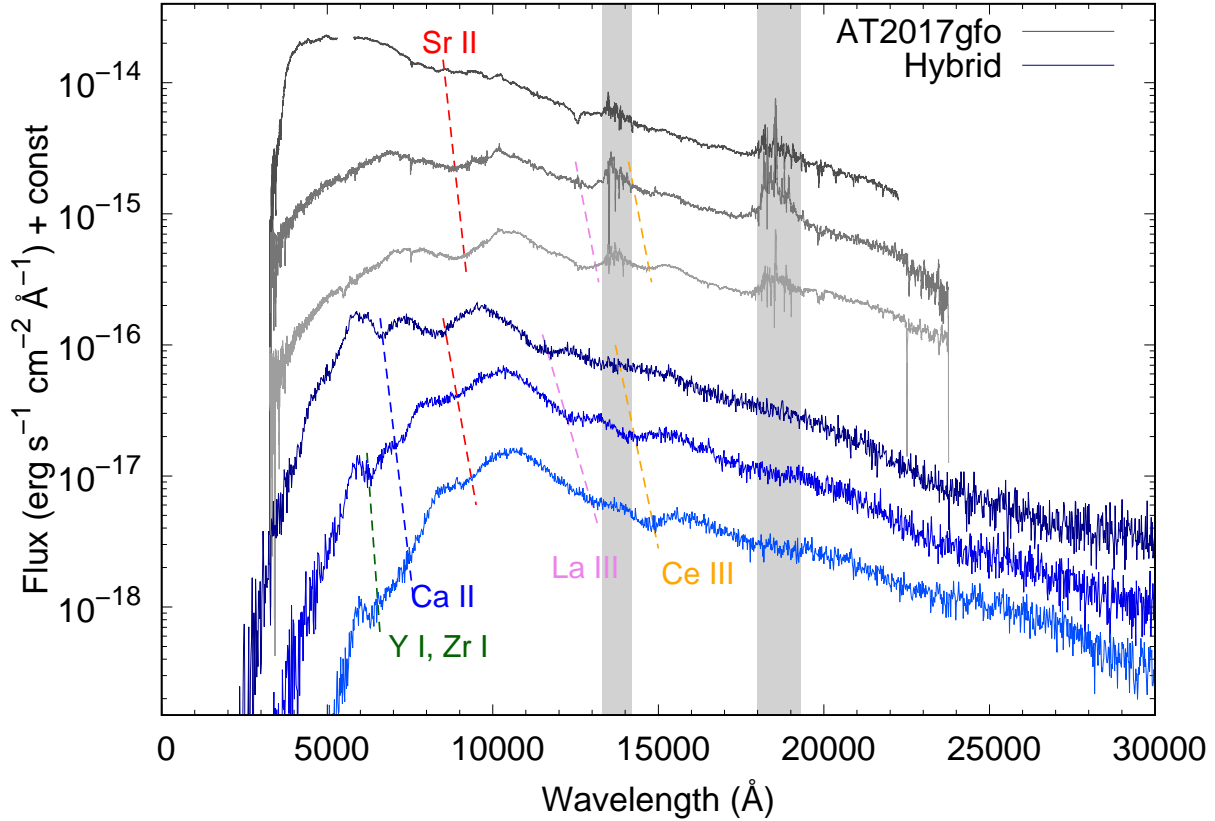}
\caption{
  \label{fig:obs}
  Comparison between the synthetic spectra (blue) and the observed spectra of AT2017gfo \citep[gray,][]{Pian2017, Smartt2017} at $t=1.5$, 2.5, and 3.5 days after the merger (dark to light colors).
  Spectra are vertically shifted for visualization.
  Gray shade shows the regions of strong atmospheric absorption.
}
\end{center}
\end{figure*}

The spectrum shows two strong absorption features at $\lambda\sim$ 8000 {\AA} (red arrow) and $\lambda\sim$ 6500 {\AA} (blue arrow).
These are mainly caused by Sr II and Ca II as found in \citet{Domoto2021}, although some Y II (dark-blue) and Zr II (green) lines also slightly affect the feature.
In addition, the spectrum shows wide absorption features around $\lambda\sim$ 12000 and 14000 {\AA} (pink and orange arrows).
These are produced by La III and Ce III lines, respectively.
The presence of these absorption features is reasonable, because Ca II, Sr II, La III, and Ce III are found to be strong absorption sources in the one-zone analysis (Section \ref{sec:linelist}).
The central wavelengths of absorption lines for La III and Ce III show that the photospheric velocity at the NIR region is $v\sim0.16\ c$, while that for Ca II and Sr II is $v\sim0.2\ c$.
This indicates that the line forming regions for different wavelength ranges do not coincide owing to the wavelength-dependent opacity.

Figure \ref{fig:obs} shows comparison between our results and the observed spectra of AT2017gfo at $t=$ 1.5, 2.5, and 3.5 days after the merger (\citealp{Pian2017, Smartt2017}; see \citealp{Gillanders2022} for the latest calibration).
We here focus only on the spectral features in the NIR region, because the absorption features by Ca and Sr in the observed spectra of AT2017gfo and their implication to the ejecta condition have already been discussed in \citet{Domoto2021}.
We find that overall slopes of synthetic spectra in the NIR region are quite similar to the observed ones.
The Doppler shift of absorption lines becomes smaller with time, because the density of the ejecta becomes lower and the photosphere moves inward (in mass coordinate).
Interestingly, the positions of absorption features at the NIR wavelengths in our results are consistent with those seen in AT2017gfo, especially at $t \ge$ 2.5 days.
Although this model motivated by the observed luminosity of AT2017gfo is quite simple, the NIR features appear to agree with the observed ones without an adjustment of, \eg density distribution.
This implies that the absorption features at the NIR wavelengths in the spectra of AT2017gfo may be caused by the La III and Ce III lines.

\begin{figure*}[th]
  \begin{center}
    \includegraphics[width=0.48\linewidth]{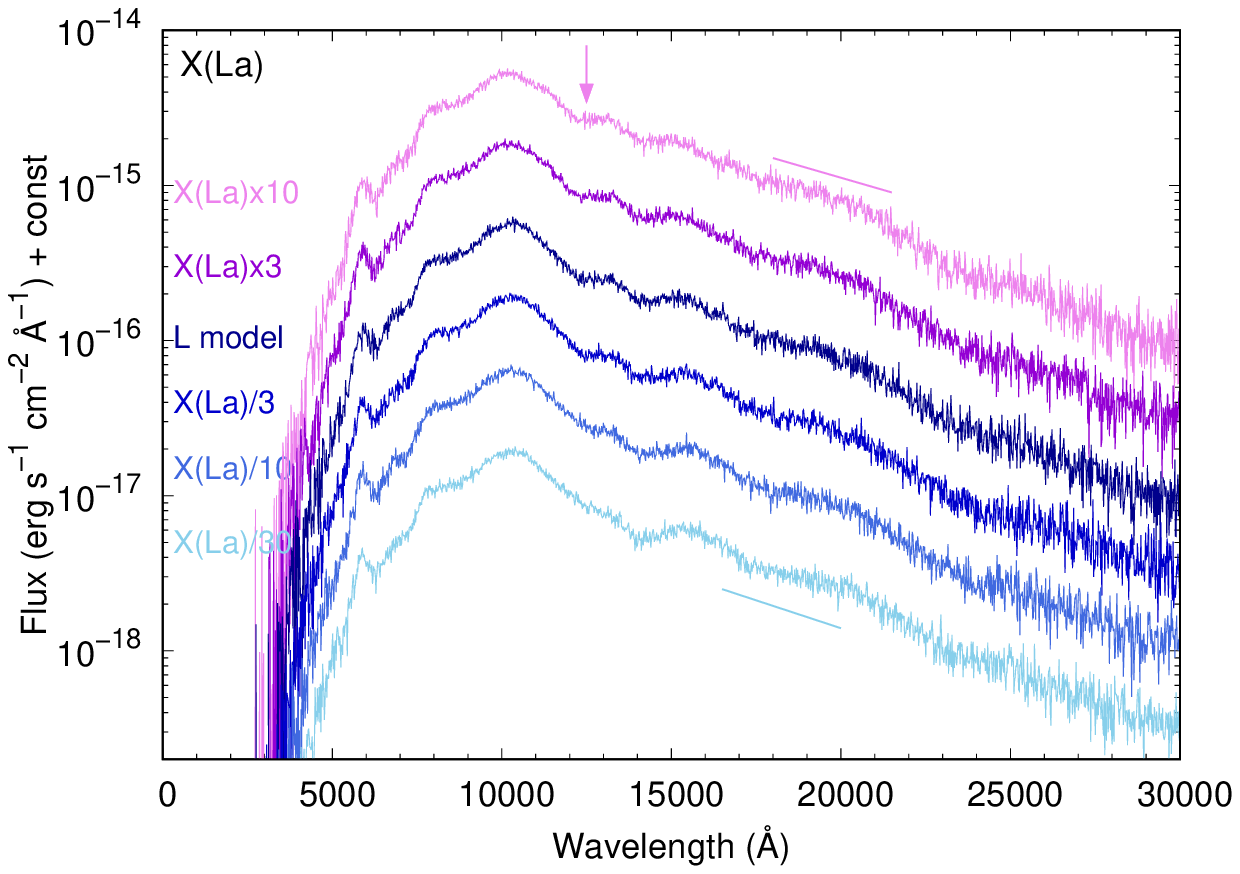}
    \includegraphics[width=0.48\linewidth]{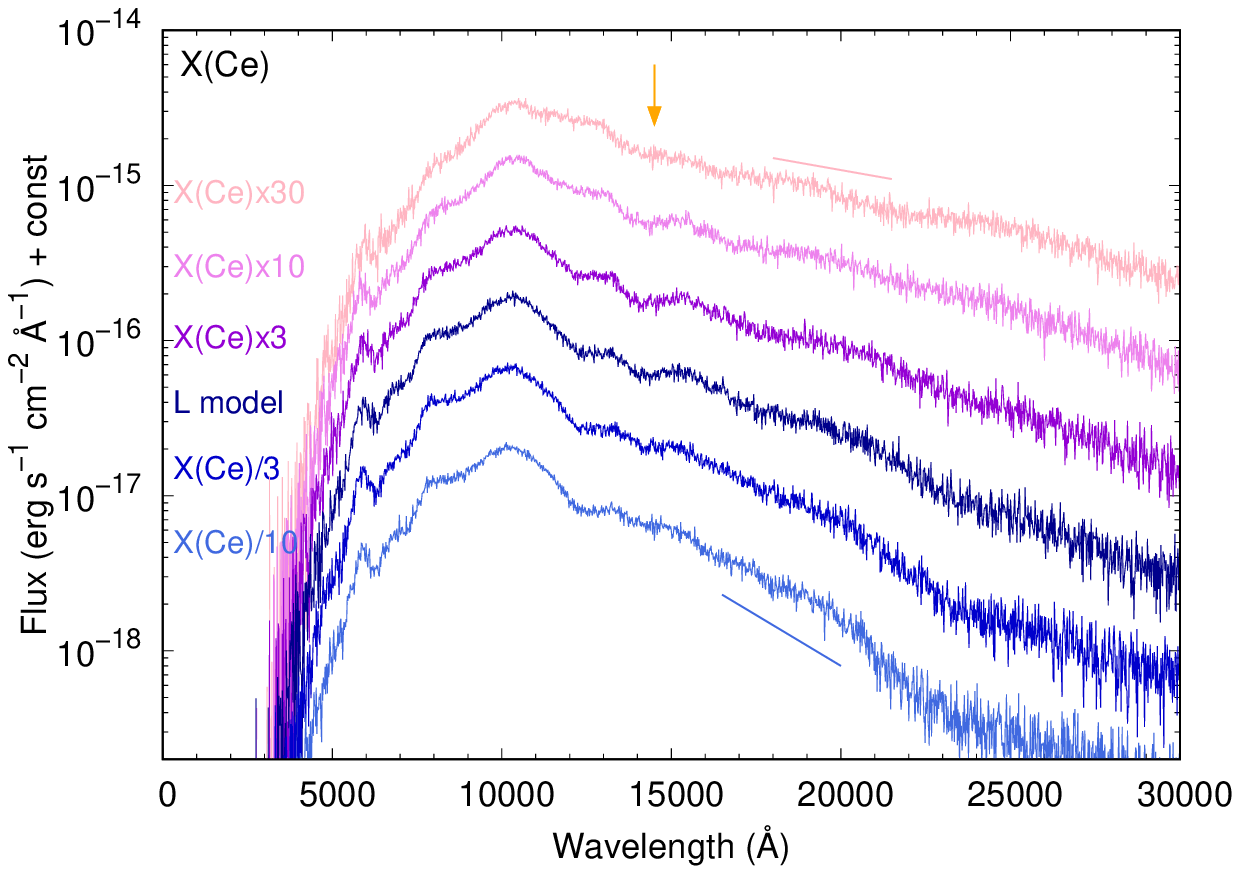}
\caption{
  \label{fig:spec lan}
  Synthetic spectra at $t=2.5$ days after the merger for different mass fractions of La (left) and Ce (right).
  Variation of each element is shown in the legend with the same color used for the spectra.
  Pink and orange arrows in each panel indicate the position of the notable absorption lines caused by La III and Ce III, respectively.
  Line segments above and below the spectra indicate spectral slopes for visualization.
}
\end{center}
\end{figure*}

It should be noted that the assumption of LTE may not be valid in a low density region.
In the results here, neutral atoms especially for Y and Zr, which appear in the outer ejecta, are the dominant opacity sources at $t \ge$ 2.5 days at the optical wavelengths \citep{Tanaka2020, Kawaguchi2021, Gillanders2022}.
On the other hand, recent work on the nebula phase of kilonovae suggests that ionization fractions as well as the temperature structure of ejecta can be deviated from those expected in LTE with time, \ie as the ejecta density decreases \citep{Hotokezaka2021, Pognan2022b}.
These non-LTE effects may change the emergent spectra at a few days after the merger mainly at the optical wavelengths, where many strong lines of neutral atoms exist \citep{Kawaguchi2021}.
Nevertheless, since the photosphere for the NIR region is located at inner ejecta where the density is enough high, non-LTE effects are expected to be subdominant \citep{Pognan2022a}.

\section{Discussion}
\label{sec:discussion}
\subsection{Lanthanide abundances}
\label{sec:lan}
Our results show that kilonova photospheric spectra exhibit absorption features of La III and Ce III in the NIR region, which are in fact similar to those seen in the spectra of AT2017gfo.
In this subsection, we  examine a possible range of these lanthanide mass fractions in the ejecta of AT2017gfo by using the NIR features.

To investigate the effect of the La amount on the spectra, we perform the same simulations as in Section \ref{sec:spectra} but by varying the mass fraction of La.
The resultant spectra at $t=2.5$ days after the merger are shown in the left panel of Figure \ref{fig:spec lan}.
We find that the strength of absorption due to the La III lines at $\lambda\sim 12500$ {\AA} changes with the mass fraction of La.
On the other hand, no matter how the mass fraction changes, the overall spectral shapes hardly change.
Because La lines have little effect on the total opacity, the NIR opacity is almost unchanged.
Thus, the strong lines of La III keep producing strong absorption as long as an enough amount of La is present.
According to the tests shown in the left panel of Figure \ref{fig:spec lan}, we estimate that the mass fraction of La is higher than 1/30 times that of the L model, \ie $X$(La) $> 2\times10^{-6}$, which is required to identify the visible absorption feature at $\lambda\sim 12500$ {\AA} in the spectra of AT2017gfo.

By contrast, the situation is different for the case of Ce.
To investigate the effect of the Ce amount on the spectra, we perform the same calculations for Ce as done for La above.
The resultant spectra at $t=2.5$ days after the merger are shown in the right panel of Figure \ref{fig:spec lan}.
We find that the absorption feature at $\lambda\sim 14000$ {\AA} diminishes as the mass fraction of Ce is substantially reduced (blueish curves).
Also, the absorption feature disappears as well even when the mass fraction of Ce is substantially increased (pink curve).
Because Ce lines appreciably contribute to the total opacity, the higher Ce mass fraction results in the higher total opacity.
As a result, the photosphere shifts outward compared to that in the L model.
This makes the photospheric temperature lower, and thus, the Ce III lines disappear.

The amount of Ce affects not only absorption features but also overall spectral shapes.
The spectra become redder and bluer when $X$(Ce) is increased and reduced, respectively.
This is because Ce has high opacity in the NIR region and the most dominant opacity source at the NIR wavelengths in this model.
As contribution of other heavy elements to the total opacity is subdominant, even if the mass fractions of all the elements with mass number larger than 100 are varied by a factor of 10, the results are almost same as those in the right panel of Figure \ref{fig:spec lan}.
Note that the element species that dominate the opacity depend on the ejecta conditions (such as density, temperature, and epoch), but generally lanthanide elements with a small atomic number (\eg Ce and Nd) tend to have larger contribution as they have more transitions from low-lying energy levels \citep{Tanaka2020, Even2020}.

We roughly estimate the mass fraction of Ce presented in the ejecta of AT2017gfo from our calculations.
It is difficult to determine the exact amount of lanthanides from absorption features, because Ce has complex effects on spectral formation as discussed above.
Nevertheless, our demonstration suggests that a certain amount of Ce must have been present in order to explain the absorption features as well as the NIR fluxes.
However, a too large amount of Ce diminishes the absorption features.
As a result, the mass fraction of Ce is estimated to be between 1/3 and 30 times that of the L model to account for the absorption feature, \ie $X$(Ce) $\sim$ (1--100)$\times10^{-5}$.
This corresponds to the lanthanide mass fraction of $\sim$ (2--200)$\times10^{-4}$ if assuming the solar abundance pattern of $r$-process elements.
While the lanthanide mass fraction estimated here is consistent with or somewhat higher than the values previously suggested \citep[e.g.,][]{McCully2017, Nicholl2017, Gillanders2022}, we emphasize that this is the first constraint on the lanthanide abundances using the absorption features in the spectra of AT2017gfo.

It should be noted that the results presented here are calculated with a single structure of ejecta (Section \ref{sec:spectra}).
The effects of ejecta properties, \eg mass and velocity, on the spectra should be systematically examined.
Furthermore, the spectra are calculated by assuming a simple one-dimensional morphology of ejecta with homogeneous abundance distribution.
It is important to employ more realistic models to elucidate the effects of multi-dimensional ejecta structures.
We leave such exploration to future work.

\subsection{Features of actinide elements}
\label{sec:actinide}
\begin{figure}[th]
  \begin{center}
    \includegraphics[width=0.95\linewidth]{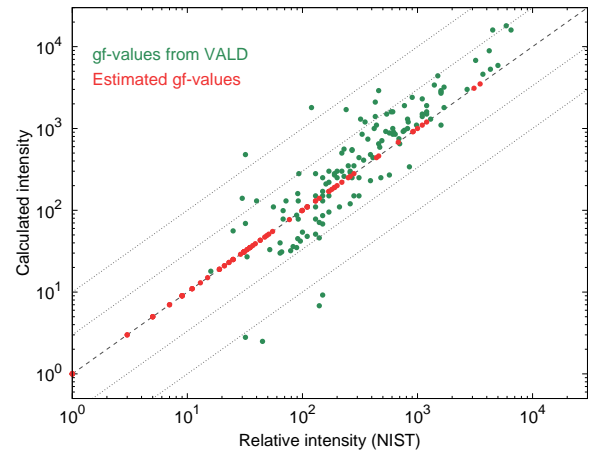}
\caption{
  \label{fig:Th3 intensity}
  Comparison of intensities (green circles) for Th III lines between those calculated with $gf$-values from the VALD database and those measured by experiments \citep{NIST_ASD, Engleman2003}.
  Gray dashed and dotted lines correspond to perfect agreement and deviations by a factor of 3 and 10, respectively.
  Red circles indicate the lines whose $gf$-values are estimated from the measured intensities.
}
\end{center}
\end{figure}

We have shown that the ions that tend to produce absorption features in kilonova photospheric spectra can be explained by atomic properties (Section \ref{sec:atomic}).
According to the required properties, one can notice that not only lanthanides but also actinide elements can possibly contribute to the spectral features.
However, while we include actinide elements in our abundance input up to $Z=100$, actinide elements are not included in the line list due to the difficulty of atomic structure calculations for actinides \citep{Tanaka2020}.
On the other hand, some experimental data are available for Th ($Z=90$).
In the following, we discuss the effects of Th absorption features based on experimental data.

\begin{figure}[th]
  \begin{center}
    \includegraphics[width=\linewidth]{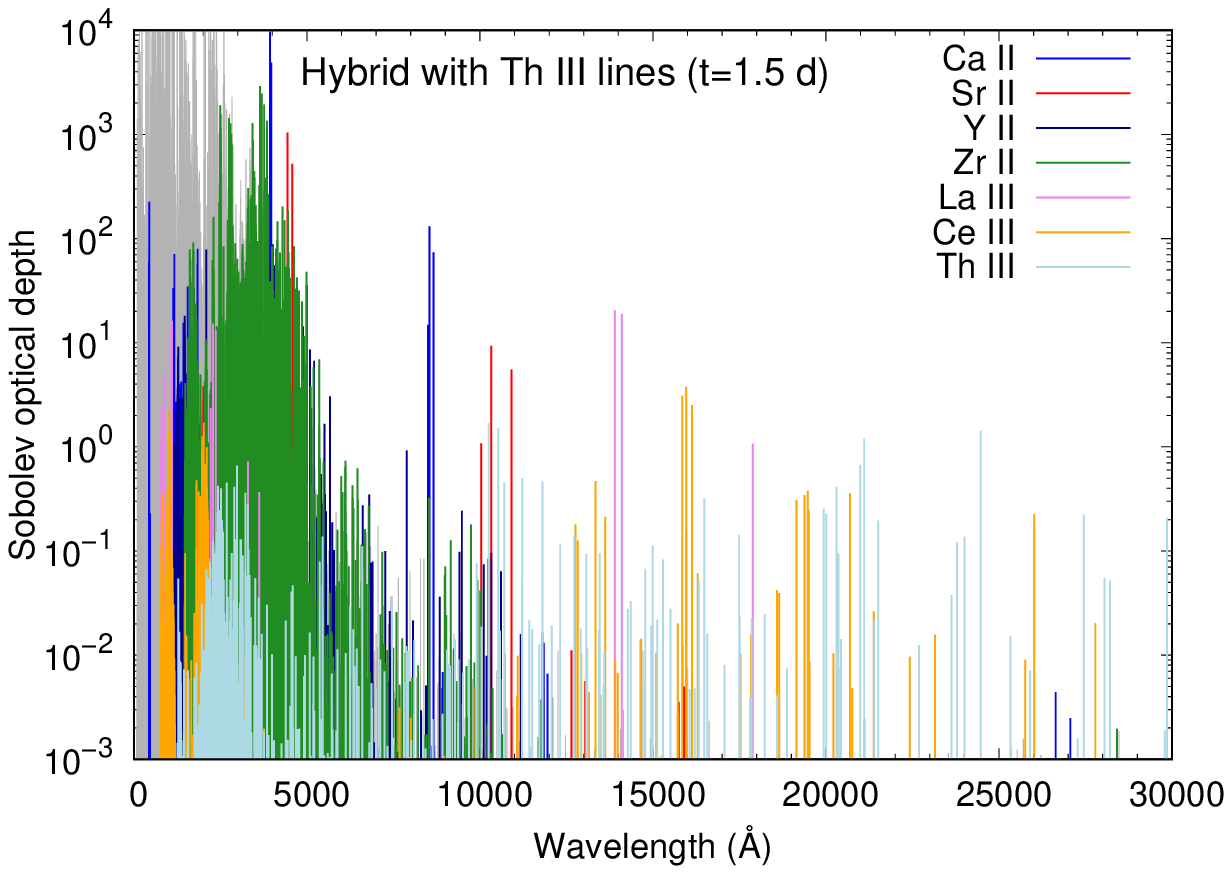} \\
    \includegraphics[width=\linewidth]{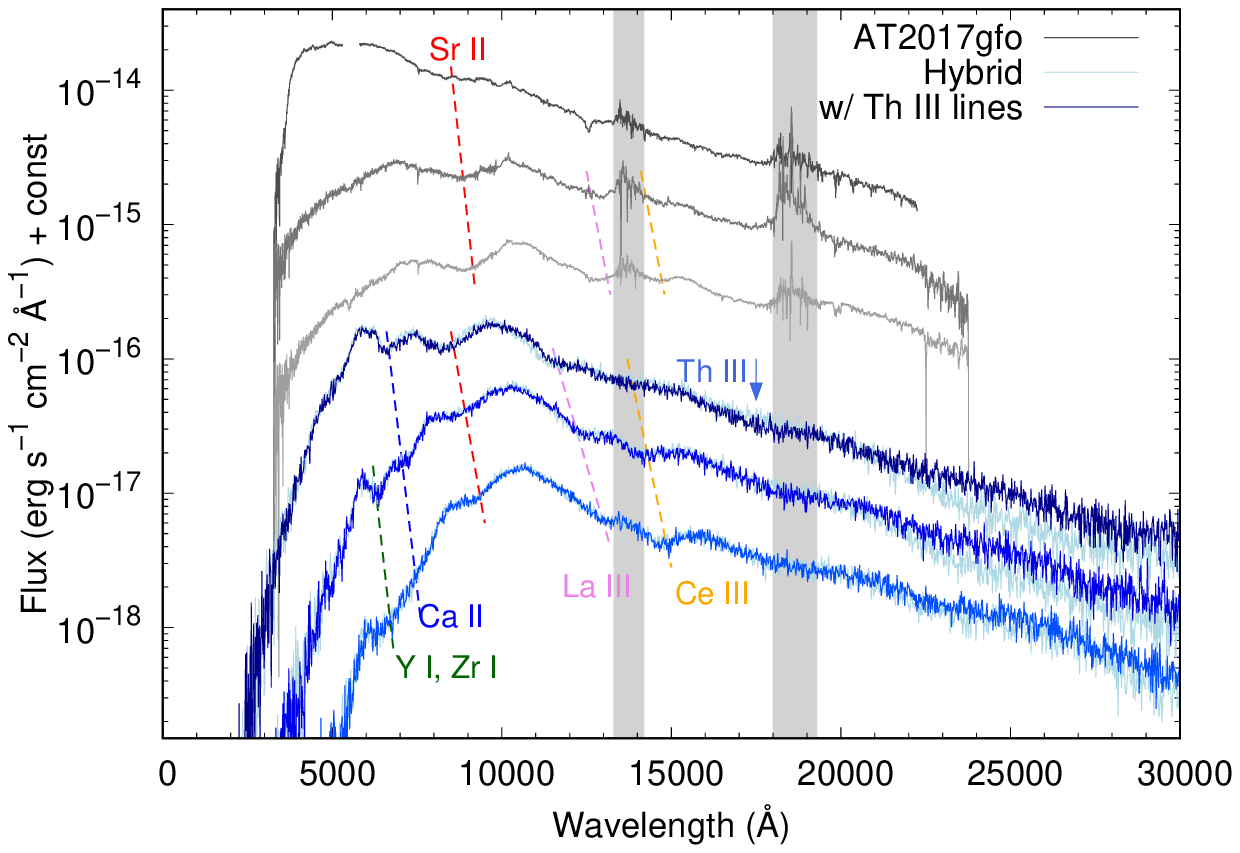}
\caption{
  \label{fig:Th3}
  Top: Sobolev optical depth of bound-bound transitions calculated with the hybrid line list (Section \ref{sec:hybrid})
  including Th III lines (light blue) taken from the VALD database ($\lambda < 10000$ {\AA}) and the NIST database ($\lambda \ge 10000$ {\AA})
  at $t=1.5$ days under the condition of $\rho=10^{-14}$ g~cm$^{-3}$ and $T=5000$ K.
  Bottom: comparison between the synthetic spectra including Th III lines (blue) and the observed spectra of AT2017gfo \citep[gray,][]{Pian2017, Smartt2017}
  at $t=1.5$, 2.5, and 3.5 days after the merger (dark to light colors).
  The light-blue curves are the same as the blue curves in Figure \ref{fig:obs}.
}
\end{center}
\end{figure}

Th III is one of the possible candidates for the ions that can contribute to the spectral features.
The atomic structure of Th III is analogous to that of Ce III that has two electrons in the outermost shell involving with $f$-shell.
Fortunately, the energy levels of Th III are well established by experiments.
For optical lines, not only transition wavelengths but also transition probabilities are available in the VALD database \citep{Biemont2002}.
Moreover, the transition wavelengths in the NIR region are measured by experiments \citep{Engleman2003}.
However, there is no available data of transition probability for the NIR lines.

To test the possibility of identifying Th III lines in kilonova photospheric spectra, we estimate the transition probabilities of the NIR Th III lines by using the measured intensities.
Since the relative intensities of the measured NIR lines \citep{Engleman2003} are listed in the NIST database \citep{NIST_ASD} in a consistent way with those of optical lines, we can directly compare the measured and calculated intensities for the lines over the whole wavelength range.
Here, although the same estimate of $gf$-values can be, in principle, applicable to other ionization stages of Th, \eg Th II, only those of Th III are tested.
This is because lines of lower-ionized ions for Th are not expected to show strong transitions compared to those of Th III, such that Ce III show strong lines but Ce II does not (Section \ref{sec:atomic}).

First, we calculate the line intensities for the optical lines for Th III with known $gf$-values taken from the VALD database using Equation \ref{eq:I}.
Comparison between the intensities calculated with the VALD $gf$-values and those measured by experiments for Th III \citep{NIST_ASD} is shown in Figure \ref{fig:Th3 intensity}.
We find that the calculated and experimentally measured intensities reasonably agree with each other when we adopt the temperature of $T=5000$ K.
Therefore, we estimate the transition probabilities of NIR lines for Th III by using the measured intensities \citep{NIST_ASD, Engleman2003} and this temperature.
Estimated $gf$-values are summarized in Table \ref{tab:Th3} (Appendix \ref{sec:Th3}).

Then, we calculate the strength of bound-bound transitions for Th III lines at $t=1.5$ days for the L model, as in Section \ref{sec:linelist}.
The top panel of Figure \ref{fig:Th3} is the same as the left bottom panel of Figure \ref{fig:tau} but with the Th III lines.
We find that the strength of the Th III lines at the NIR wavelengths can be comparable to that of Ce III lines.
This is due to the same reason as discussed in Section \ref{sec:atomic}: relatively high transition probabilities and low energy levels of transitions.

We perform the same radiative transfer simulations as in Section \ref{sec:spectra} by including the Th III lines.
The synthetic spectra with the Th III lines are shown in the bottom panel of Figure \ref{fig:Th3} (blue curves).
We find that the fluxes for $\lambda \ge 20000$ {\AA} are slightly pushed up compared to those of the results without the Th III lines (light-blue curves, see Figure \ref{fig:obs}).
However, the spectral features are not substantially different from those not including the Th III lines, although a wide and marginal absorption feature can be seen around $\lambda\sim 18000$ {\AA} at $t=1.5$ days.
This implies that it is difficult to confirm the presence of Th from spectral features.

The difference between the absorption features for Th III and Ce III can be explained by the complex atomic structure of actinides. 
As shown in the top panel of Figure \ref{fig:Th3}, although the Th III lines in the NIR region are relatively strong, many transitions exhibit similar Sobolev optical depths and no significant line like that of Ce III at $\lambda\sim 16000$ {\AA} exists.
This is due to the fact that Th III has denser low-lying energy levels involved in 5$f$-shell compared to those of Ce III involved in 4$f$-shell.
\citet{Silva2022} recently showed that the opacity of another actinide U III ($Z=92$) is about an order of magnitude larger than that of Nd III for the same reason.
Note that an actinide, Ac ($Z=89$, as $^{227}$Ac with the half-life of 21.77 yr), can also exist with a similar amount to those of Th and U in the ejecta of kilonovae.
Ac III may have a similar atomic structure to that of La III.
However, Ac III is poorly understood both in theory and experiments, and it is not clear if Ac III can produce prominent features as La III.
Thus, although actinide ions can be important opacity sources in the NIR region, it may not be easy to identify the presence of actinides from the spectral features.

\section{Conclusions}
\label{sec:conclusion}
We have performed systematic calculations of the strength of bound-bound transitions and radiative transfer simulations with the aim of identifying elements in kilonova spectra.
We constructed a hybrid line list by combining an experimentally calibrated {\it accurate} line list with a theoretically constructed {\it complete} line list.
This allows us to investigate the entire wavelength range of kilonova photospheric spectra.
We have found that La III and Ce III produce absorption features at the NIR wavelengths ($\lambda\sim 12000$--14000 {\AA}).
The positions of these features are consistent with those seen in the spectra of GW170817/AT2017gfo.
Using the absorption lines caused by La III and Ce III, we have estimated that the mass fractions of La and Ce synthesized in the ejecta of GW170817/AT2017gfo are $X$(La) $> 2\times10^{-6}$ and $X$(Ce) $\sim$ (1--100)$\times10^{-5}$, respectively.
This is the first spectroscopic estimation of the lanthanide abundances in NS merger ejecta.

We have shown that the elements on the left side of the periodic table (Ca, Sr, Y, Zr, Ba, La, and Ce) tend to produce prominent absorption features in kilonova photospheric spectra.
This is due to the fact that such ions have a relatively small number of valence electrons in the outermost shell (leading to low complexities, and high transition probabilities for bound-bound transitions) and have relatively low-lying energy levels (leading to a large population in the Boltzmann distribution).

Since the atomic structure of Th III is analogous to that of Ce III, we have investigated the possibility of identifying Th III lines in kilonova spectra.
We have found that it is more difficult to identify the definitive features caused by Th III, because it has denser low-lying energy levels and no outstanding identifiable lines.
Although the atomic data of Th III (Table \ref{tab:Th3}) are still uncertain, our demonstration suggests that we need to consider another way to obtain evidence of synthesized actinide elements from observables.

In this paper, we have used a model dominated by relatively light $r$-process elements.
For more lanthanide-rich ejecta, the emergent spectra should be redder and fainter than predicted by our results, due to the high opacity of heavy elements.
Also, the spectra should become smoother due to the presence of many weak lines from heavy elements.
Since our hybrid line list is constructed assuming an abundance model dominated by light $r$-process elements, we are unable to discuss spectral features in the lanthanide-rich ejecta.
To extract more information from various spectra, further effort is necessary to construct spectroscopically accurate atomic data for heavy elements.
It is also cautioned that the abundance with $Z\le19$ have been excluded in our calculations, because their mass fractions are very small ($<10^{-4}$) except for He (the left panel of Figure \ref{fig:abun}).
Although He might produce absorption features in spectra, it requires the consideration of non-LTE effects to the level population \citep{Perego2022}.
While the exclusion of light elements does not affect our conclusions under the assumption of LTE, the systematic exploration of He line formation will be of interest in future.

\begin{acknowledgments}
Part of numerical simulations presented in this paper were carried out on Cray XC50 at Center for Computational Astrophysics, National Astronomical Observatory of Japan.
N.D. acknowledges support from Graduate Program on Physics for the Universe (GP-PU) at Tohoku University.
This research was supported by NIFS Collaborative Research Program (NIFS22KIIF005), the Grant-in-Aid for JSPS Fellows (22J22810), the Grant-in-Aid for Scientific Research from JSPS (19H00694, 20H00158, 21H04997, 21K13912), and MEXT (17H06363).
\end{acknowledgments}

\appendix

\section{Calibration of atomic data}
\label{sec:lsj}
We calibrate the theoretical atomic data with experimental data to enable us to discuss absorption lines in kilonova spectra.
Here, we describe our method of this calibration procedure.
We perform the calibration for Sr II, Y I, Y II, Zr I, Zr II, Ba II, La III and Ce III, which are found as strong absorption sources in Section \ref{sec:candidate}.
For the experimental data, we use the NIST Atomic Spectra Database \citep{NIST_ASD} to calibrate the energy levels.
The energy levels of these ions have been well determined from experiments mainly in the optical wavelengths.

The NIST database lists term symbols of energy levels.
By using those symbols, it is possible to associate the theoretical energy levels to those in the NIST database.
It should be, however, noted that energy terms can be expressed by different ways depending on angular momentum coupling schemes: LS-coupling and jj-coupling.
While the NIST database adopts the LS scheme, the HULLAC code used for atomic calculations adopts the jj scheme \citep{HULLAC}.
Since it is not possible to directly compare energy levels in different schemes, we perform the transformation from the jj-coupled to the LS-coupled energy terms for the theoretical energy levels \citep{Cowan1968, Cowan1981}.

While energy terms are uniquely determined for one-electron systems (Sr II, Ba II, and La III), the transformations are required for atoms with more than two valence electrons.
To perform all the transformations systematically, we use the LSJ code \citep{Gaigalas2004}.
The LSJ code transforms a jj-coupled basis to a LS-coupled representation according to inputs of configuration state functions (CSF) and mixing coefficients.
For the input of the LSJ code, we prepare the CSF lists for each ion by means of GRASP2018 \citep{grasp2018} and mixing coefficients for energy levels from the HULLAC results \citep{Tanaka2020}.
Using these inputs, we perform the jj-LS transformations for the energy levels of Y I, Y II, Zr I, Zr II, and Ce III.

As an energy term for each energy level, we assign the leading term with the largest mixing coefficient from the results of the LSJ code.
When leading terms are the same for two levels, we assign the term to a level with the larger mixing coefficient than the other one.
For the other level, unassigned term with the next largest coefficient is assigned.

For the spectral features or opacity of kilonovae, calibration of low-lying energy levels is the most important (Section \ref{sec:atomic}).
Nevertheless, we also perform the transformation for many excited levels so that we have enough number of transitions to confirm the accuracy of $gf$-values (see below, the right panels of Figures \ref{fig:SrII}--\ref{fig:CeIII}).
Since ions with a smaller number of valence electrons tend to have a smaller number of levels, we perform the calibration for a larger number of configurations for simpler ions.
As a result, our calibrated line list naturally includes strong and important transitions.
The energy diagrams of calibrated configurations for each ion are shown in the left panels of Figure \ref{fig:SrII}--\ref{fig:CeIII}.

For Zr I, the calibration is performed in a slightly different way from other ions.
When angular momenta of more than three electrons are coupled, intermediate terms are needed to distinguish the energy terms.
However, for the energy terms of most levels in 4$d^2$5$s$5$p$ for Zr I, the NIST database shows the intermediate terms in a different way from the LSJ code.
Therefore, we associate the energy terms for 4$d^2$5$s$5$p$ of the HULLAC results and the NIST database in the order of energy among levels with the same total angular momentum $J$.

After calibration of the energy levels, we calibrate the wavelengths of the transitions between the calibrated energy levels.
Since we aim at discussing the spectral features of kilonovae, the lines whose original and calibrated wavelengths are in the forest of lines at $\lambda < 5000$ {\AA} are left as the original theoretical ones for simplicity.
Then, if available, transition probabilities of the calibrated lines are taken from the VALD database \citep{Piskunov1995, Kupka1999, Ryabchikova2015}.
For Zr I and Zr II, we instead use the Kurucz's atomic data\footnote{http://kurucz.harvard.edu/atoms.html} \citep{Kurucz2018}, which are constructed by semi-empirical calculations and newer than those in the VALD database.
We use these database instead of the NIST database, because the NIST database do not necessarily include all the transition data for heavy elements.
If the transition probabilities of the calibrated lines are not listed in both database, we adopt those from the theoretical calculations, which only happens to La III and Ce III mainly in NIR wavelengths.
We summarize the calibrated lines that adopt the theoretical $gf$-values with $\lambda>7000$ {\AA} and log $gf > -3$ in Table \ref{tab:LaIII}--\ref{tab:CeIII}.

Note that, theoretically calculated transition probabilities are not necessarily accurate.
The right panels of Figure \ref{fig:SrII}--\ref{fig:CeIII} show comparison of $gf$-values between the HULLAC results and the VALD (or Kurucz's) database for all the lines between the calibrated energy levels.
We see that, while the values are roughly agree for simple ions, there is a scatter as the number of outermost electrons increases, especially for low $gf$-values.
This is unavoidable, as the theoretical calculations become inaccurate as atomic structures become complex.
Nevertheless, it is emphasized that uncertainty of theoretical $gf$-values do not affect our conclusions about NIR spectral features, because $gf$-values of La III and Ce III agree quite well for strong transitions as shown in the right panel of Figure \ref{fig:LaIII} and \ref{fig:CeIII} (see also Figure \ref{fig:intensity} for the NIR Ce III lines).
To determine the exact values of transition probabilities of the lines, more experimental and observational calibrations are necessary.

\section{Estimated transition probabilities of Th III lines}
\label{sec:Th3}
We summarize the $gf$-values of Th III lines estimated by using the measured line intensities in Table \ref{tab:Th3} (see Section \ref{sec:actinide}).

\begin{figure*}[th]
  \begin{center}
    \begin{tabular}{cc}
    \includegraphics[width=0.51\linewidth]{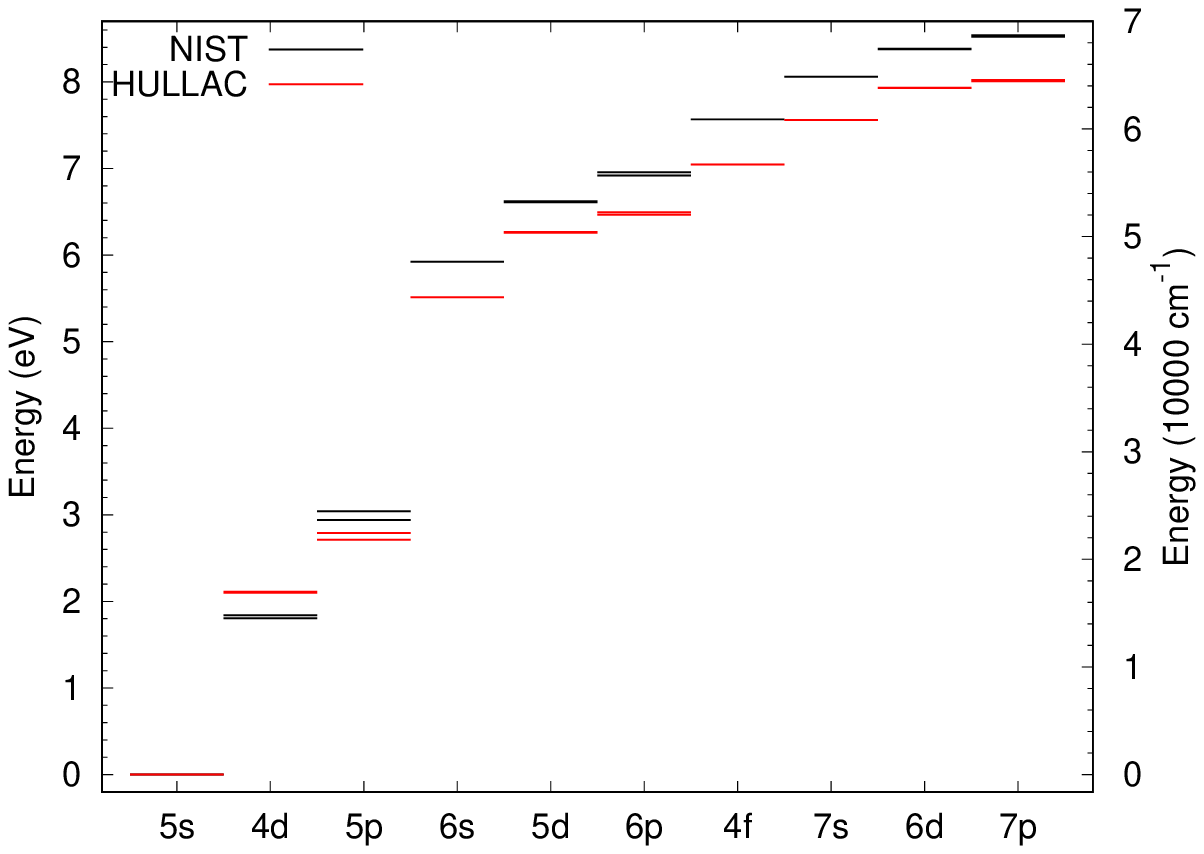} &
    \includegraphics[width=0.44\linewidth]{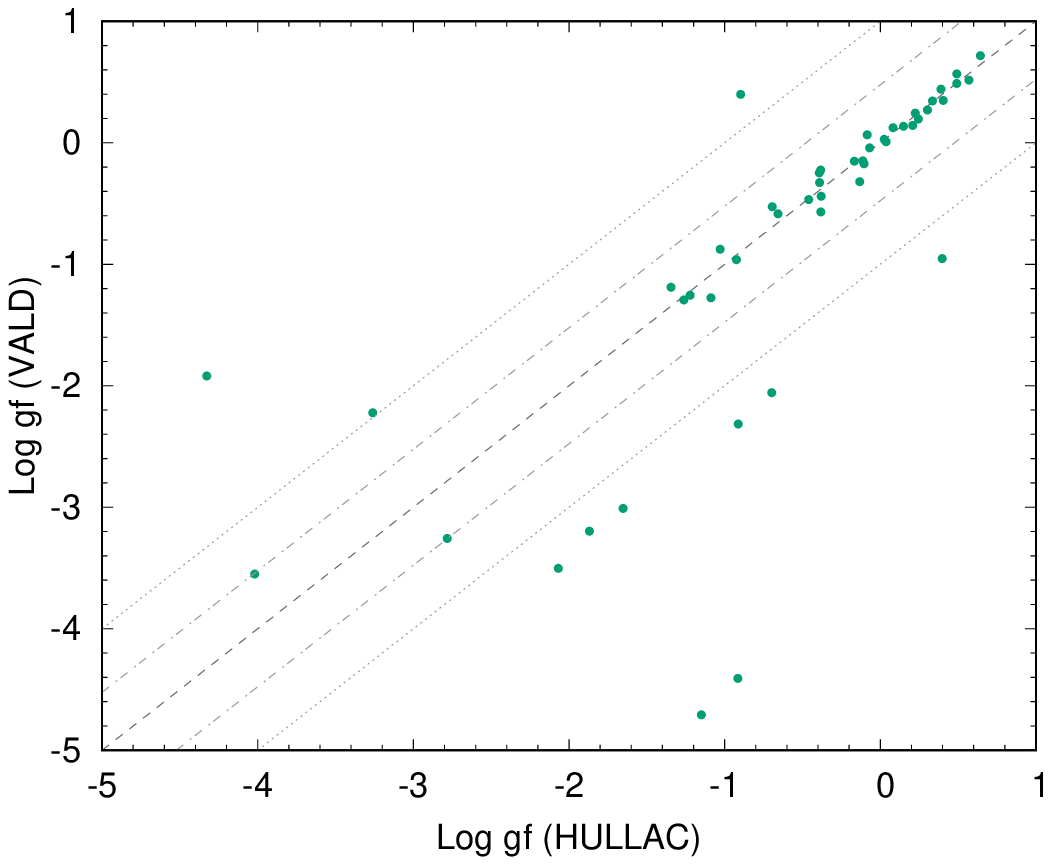}  \\
    \end{tabular}
\caption{
  \label{fig:SrII}
  Left: energy diagram for Sr II. Black and red lines show energy levels from the NIST database and the HULLAC results, respectively.
  Right: comparison of $gf$-values between the VALD database and the HULLAC results.
  Gray dashed line corresponds to perfect agreement between them, and dashdotted and dotted lines indicate deviations by a factor of 3 and 10, respectively.
}
\end{center}
\end{figure*}
\begin{figure*}[th]
  \begin{center}
    \begin{tabular}{cc}
    \includegraphics[width=0.51\linewidth]{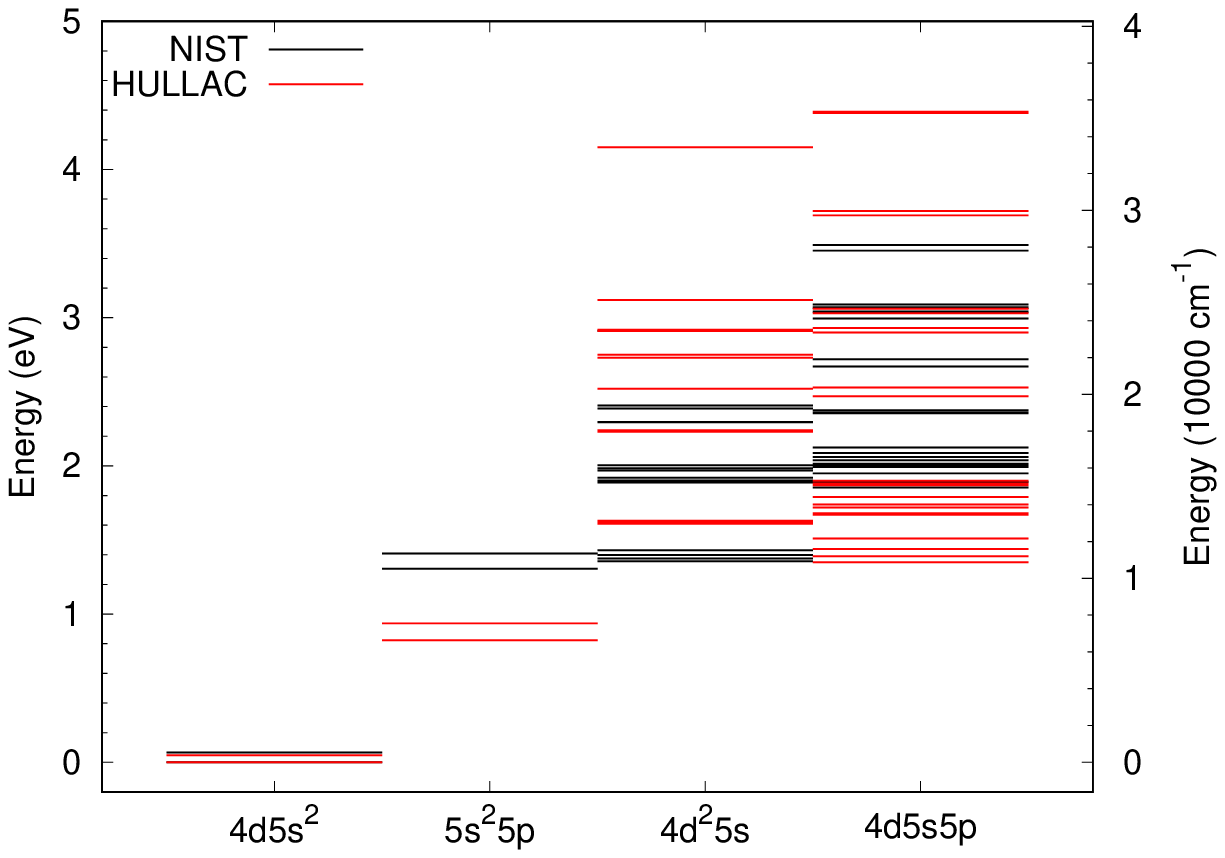} &
    \includegraphics[width=0.44\linewidth]{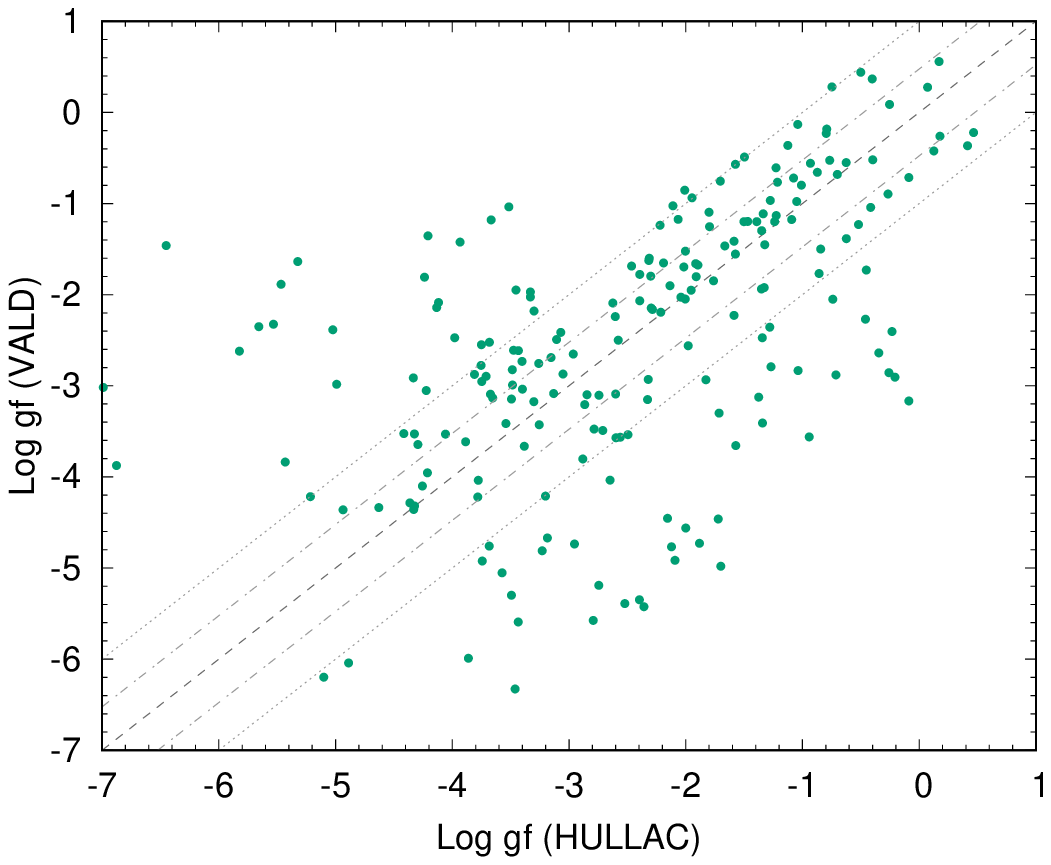}  \\
    \end{tabular}
\caption{
  \label{fig:YI}
  Same as Figure \ref{fig:SrII}, but for Y I.
}
\end{center}
\end{figure*}
\begin{figure*}[th]
  \begin{center}
    \begin{tabular}{cc}
    \includegraphics[width=0.51\linewidth]{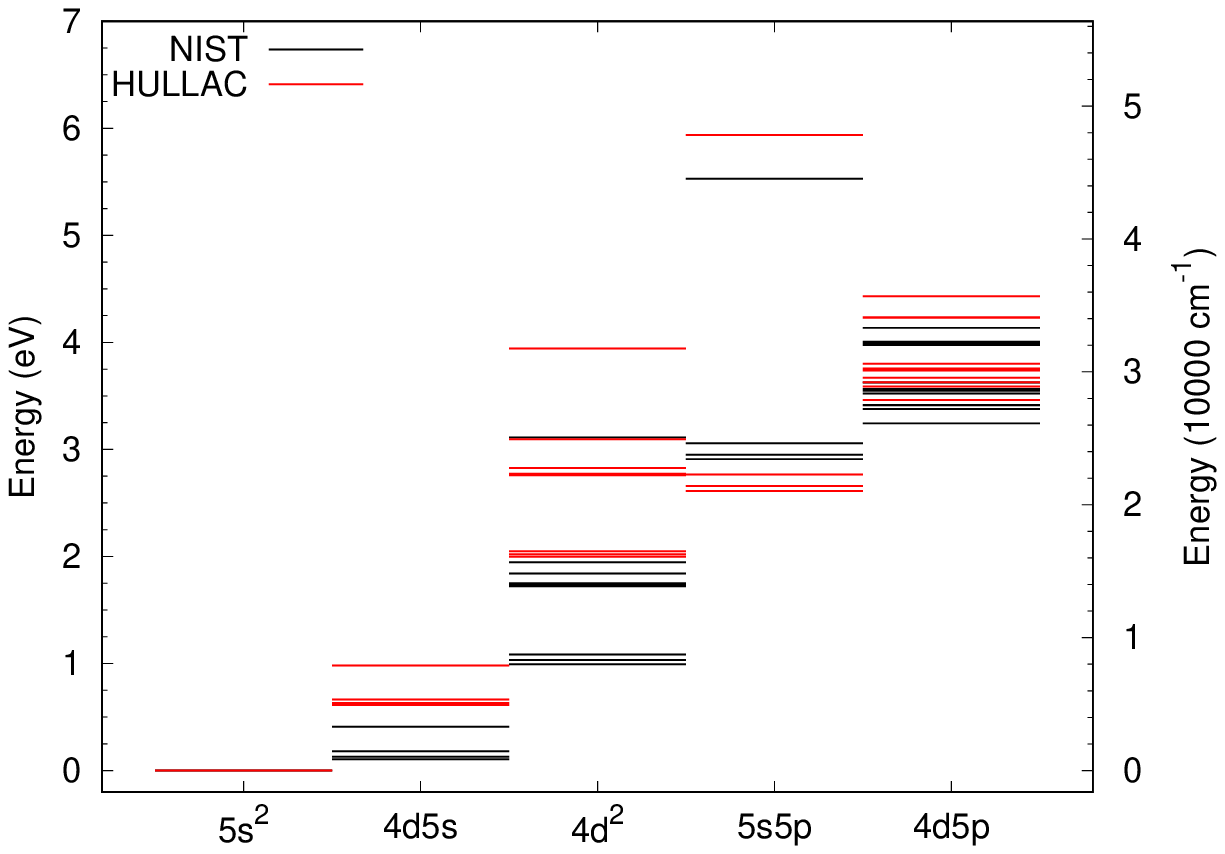} &
    \includegraphics[width=0.44\linewidth]{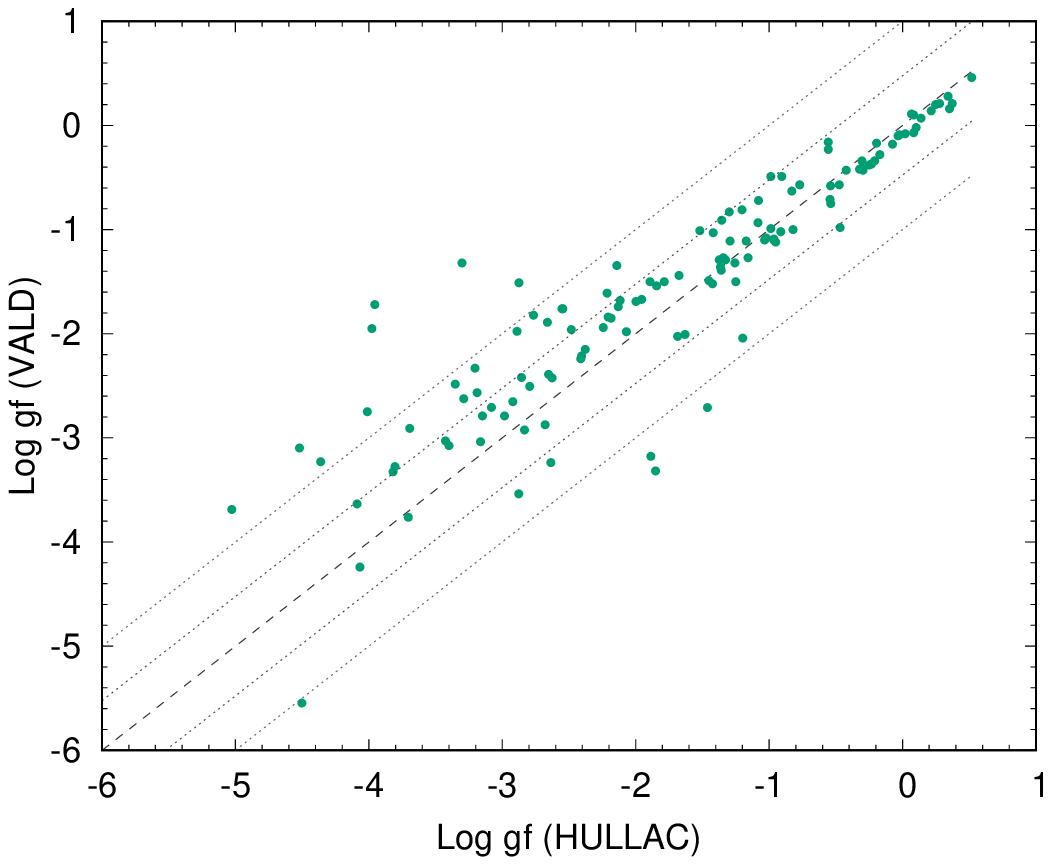}  \\
    \end{tabular}
\caption{
  \label{fig:YII}
  Same as Figure \ref{fig:SrII}, but for Y II.
}
\end{center}
\end{figure*}
\begin{figure*}[th]
  \begin{center}
    \begin{tabular}{cc}
    \includegraphics[width=0.51\linewidth]{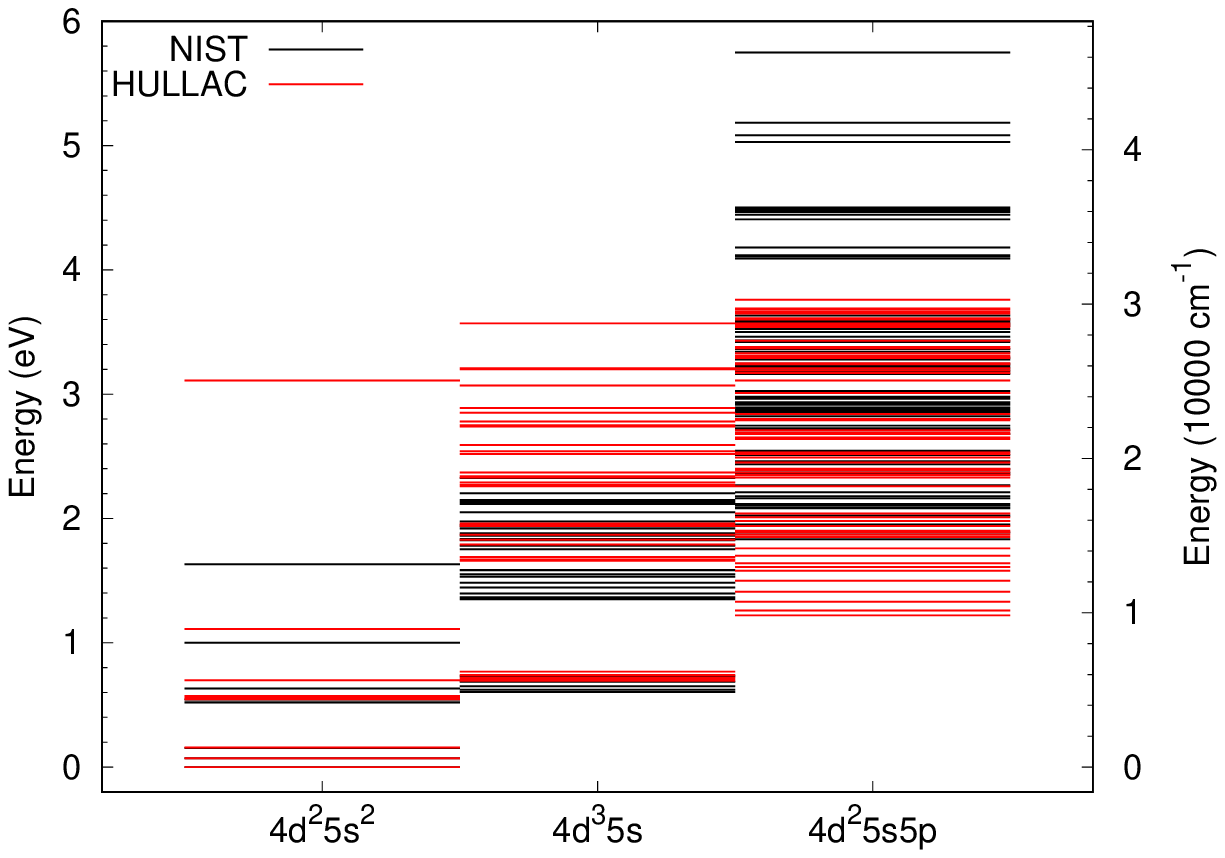} &
    \includegraphics[width=0.44\linewidth]{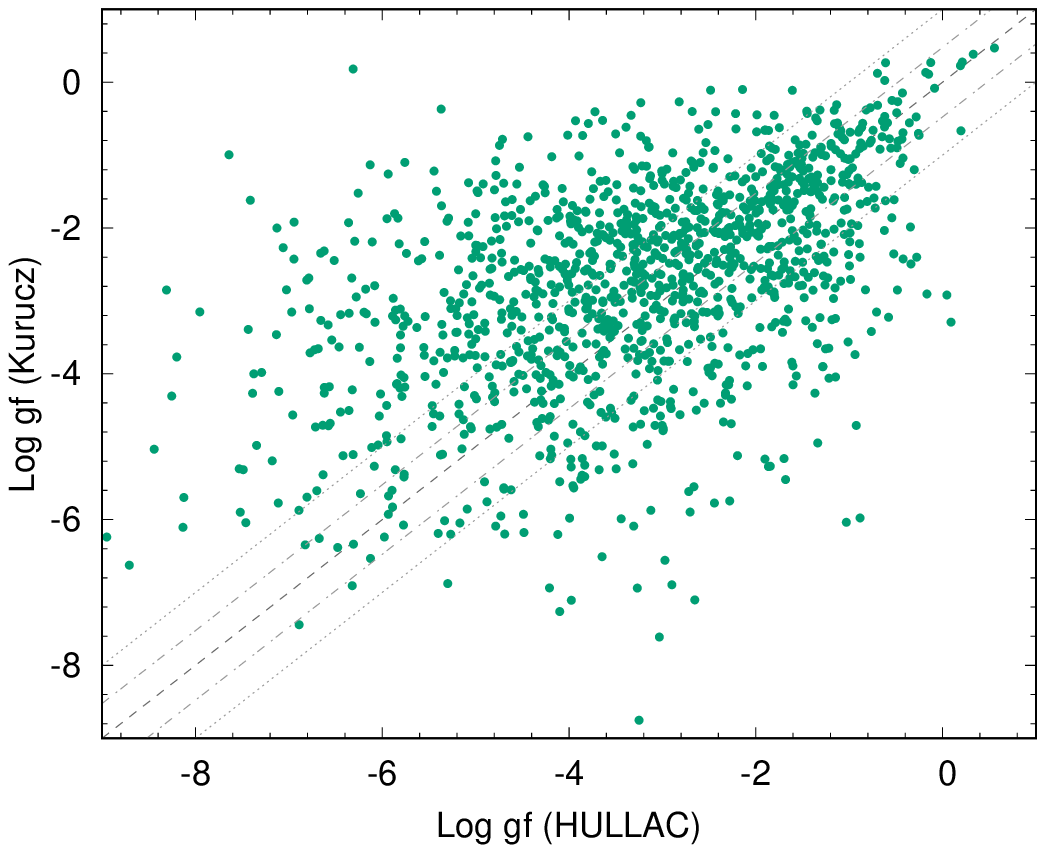}  \\
    \end{tabular}
\caption{
  \label{fig:ZrI}
  Same as Figure \ref{fig:SrII}, but for Zr I.
  Kurucz's atomic data are used instead of the VALD database for comparison in the right panel.
  The energy levels of 4$d^2$5$s$5$p$ above 4 eV for the HULLAC results are not shown 
  and not used for the calibration because of strong mixing with high-lying levels of other configurations.
}
\end{center}
\end{figure*}
\begin{figure*}[th]
  \begin{center}
    \begin{tabular}{cc}
    \includegraphics[width=0.51\linewidth]{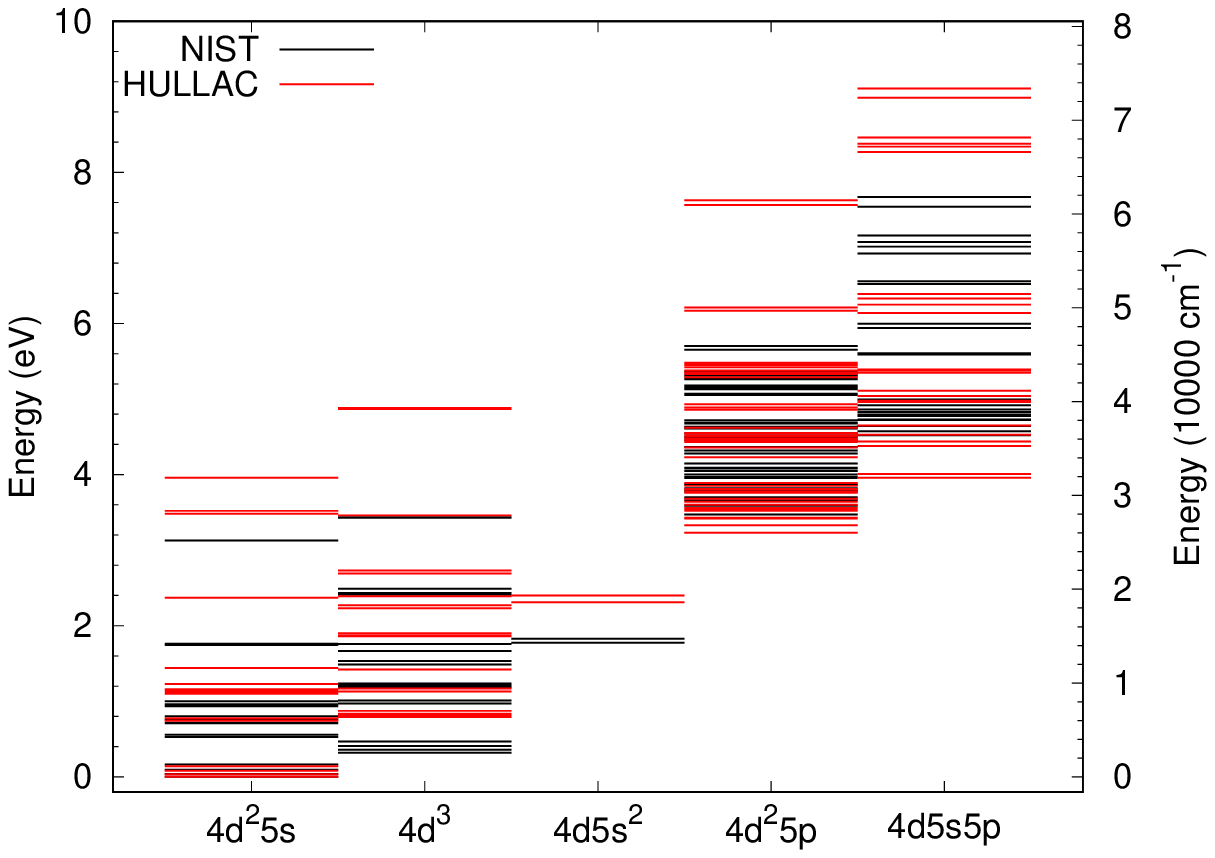} &
    \includegraphics[width=0.44\linewidth]{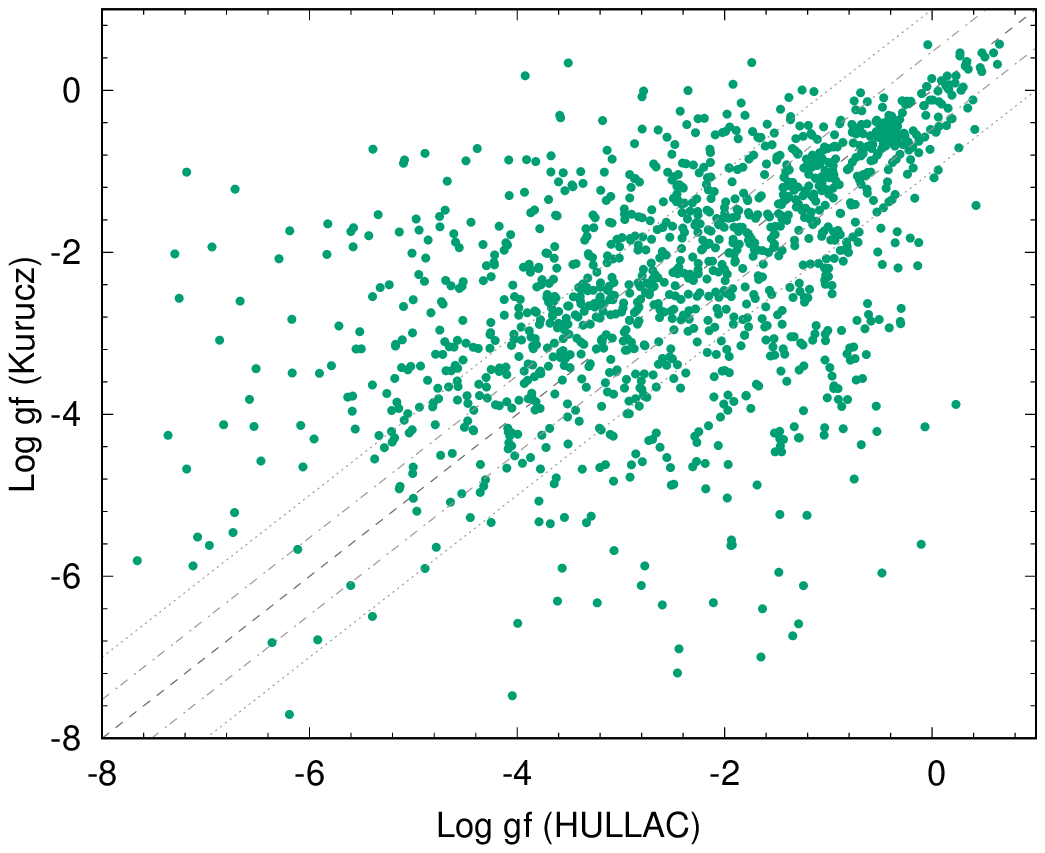}  \\
    \end{tabular}
\caption{
  \label{fig:ZrII}
  Same as Figure \ref{fig:SrII}, but for Zr II.
  Kurucz's atomic data are used instead of the VALD database for comparison in the right panel.
}
\end{center}
\end{figure*}
\begin{figure*}[th]
  \begin{center}
    \begin{tabular}{cc}
    \includegraphics[width=0.51\linewidth]{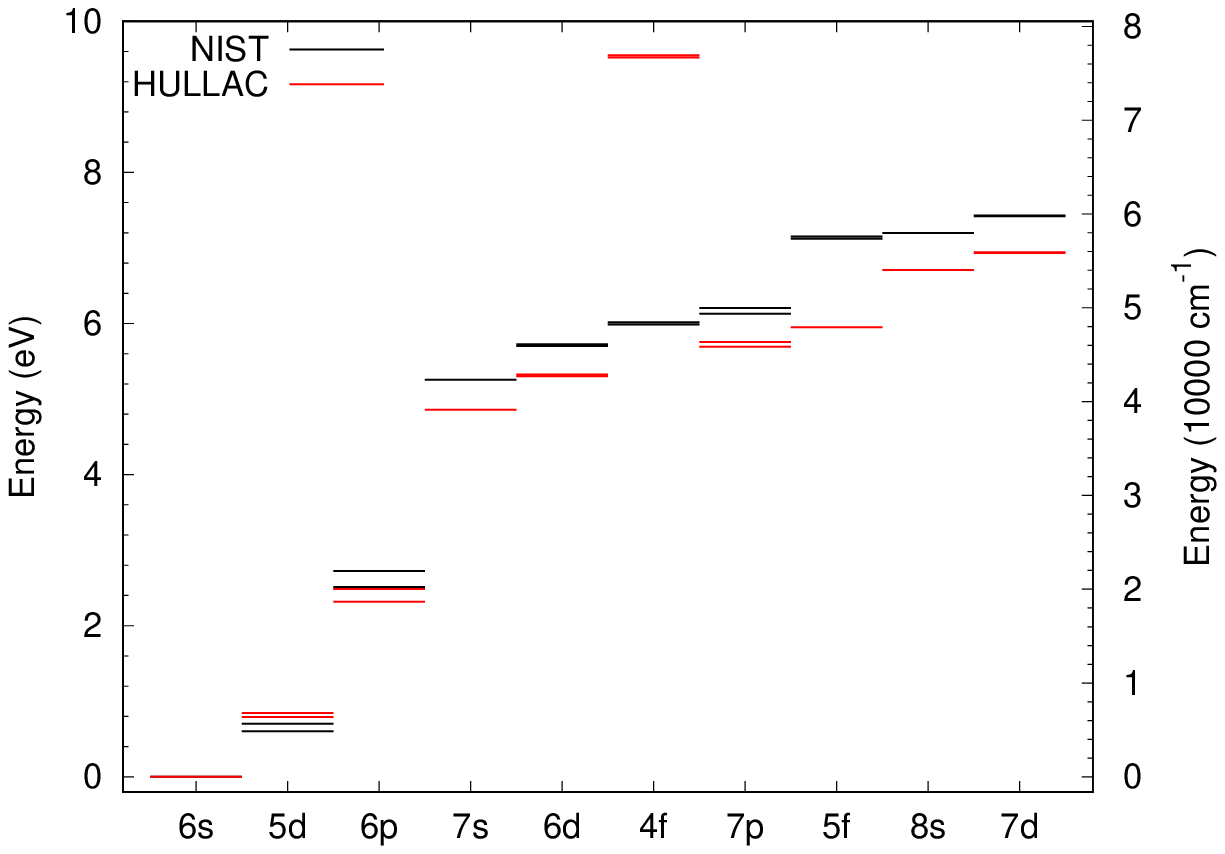} &
    \includegraphics[width=0.44\linewidth]{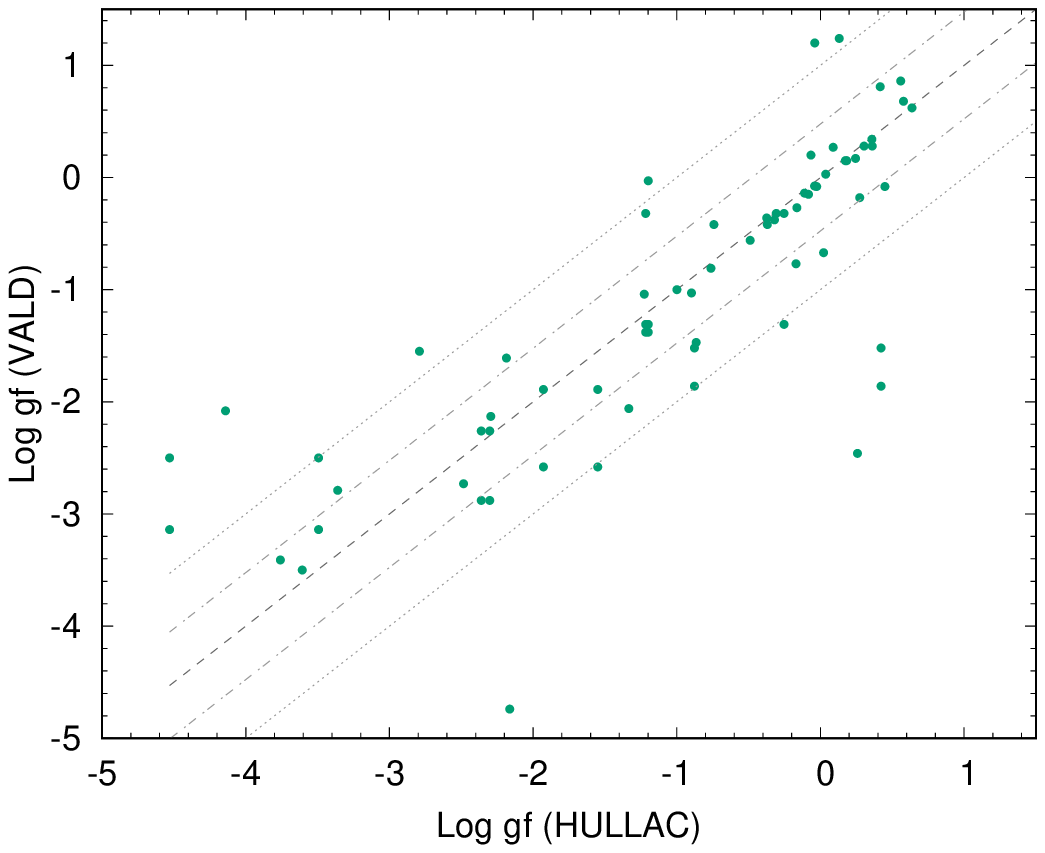}  \\
    \end{tabular}
\caption{
  \label{fig:BaII}
  Same as Figure \ref{fig:SrII}, but for Ba II.
}
\end{center}
\end{figure*}
\begin{figure*}[th]
  \begin{center}
    \begin{tabular}{cc}
    \includegraphics[width=0.51\linewidth]{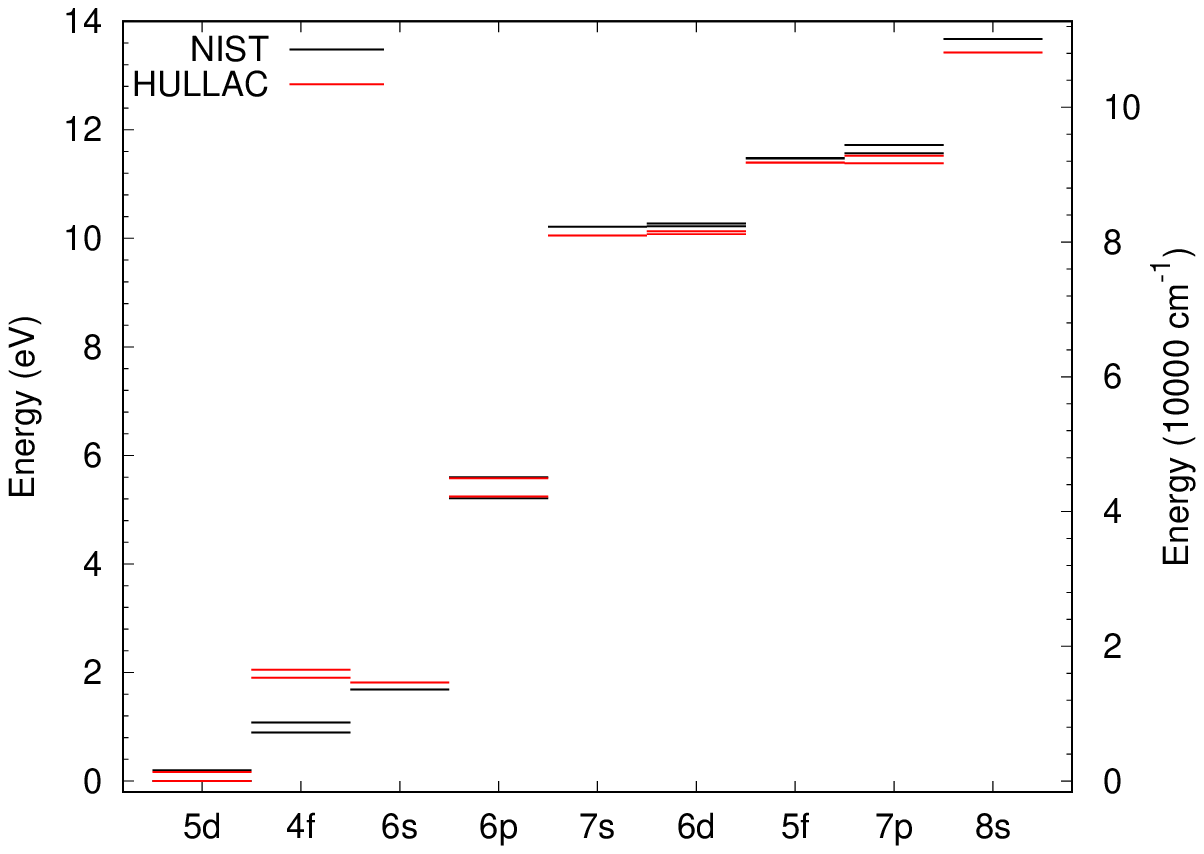} &
    \includegraphics[width=0.44\linewidth]{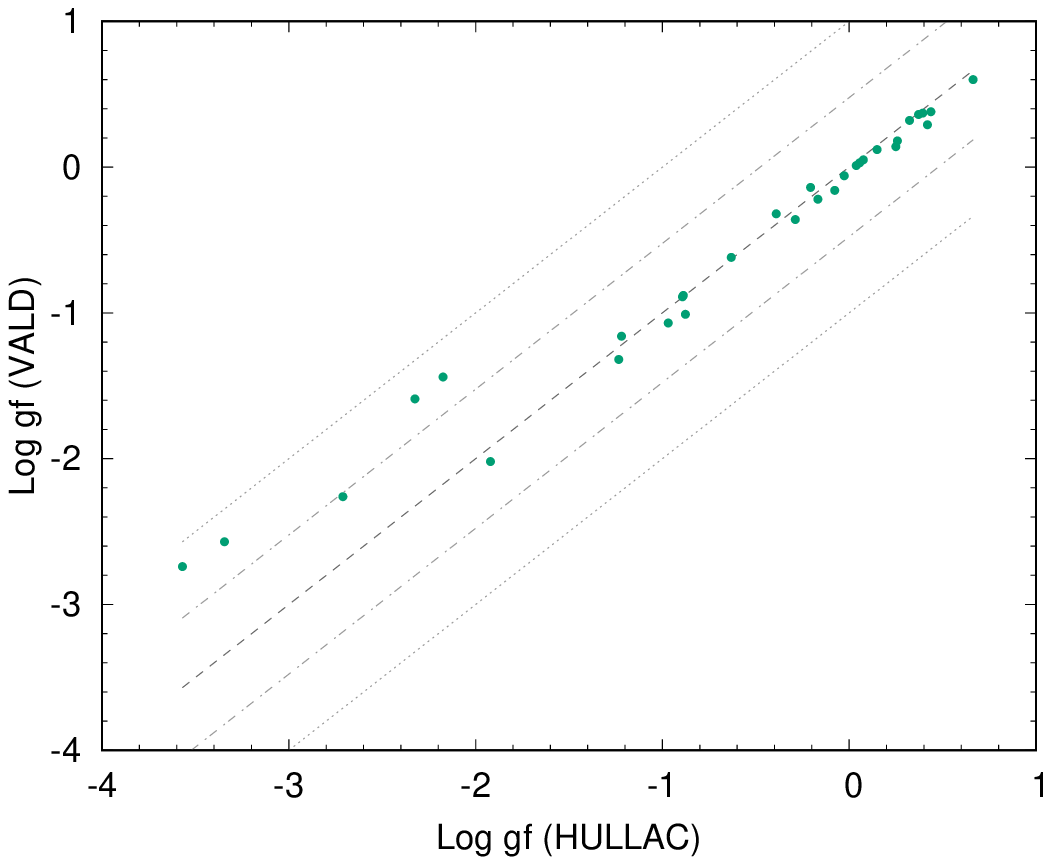}  \\
    \end{tabular}
\caption{
  \label{fig:LaIII}
  Same as Figure \ref{fig:SrII}, but for La III.
}
\end{center}
\end{figure*}
\begin{figure*}[th]
  \begin{center}
    \begin{tabular}{cc}
    \includegraphics[width=0.51\linewidth]{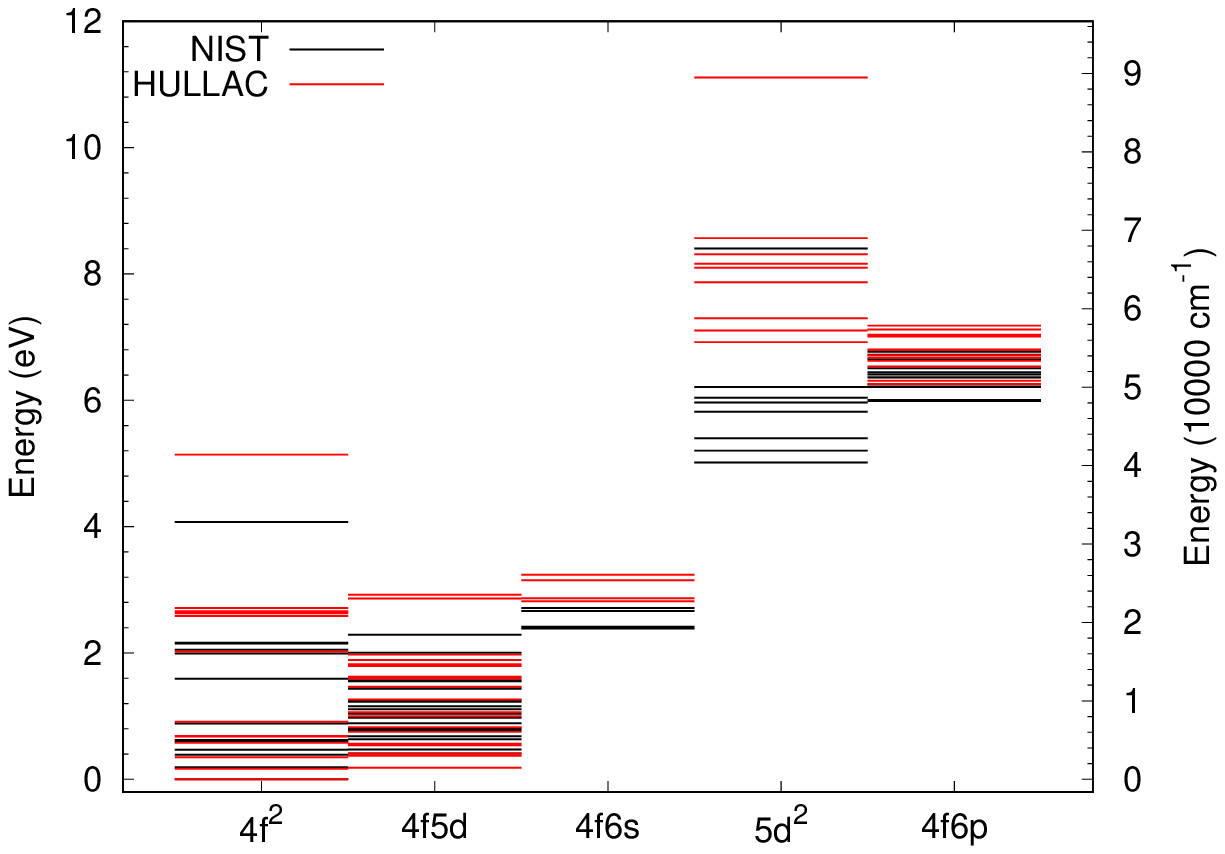} &
    \includegraphics[width=0.44\linewidth]{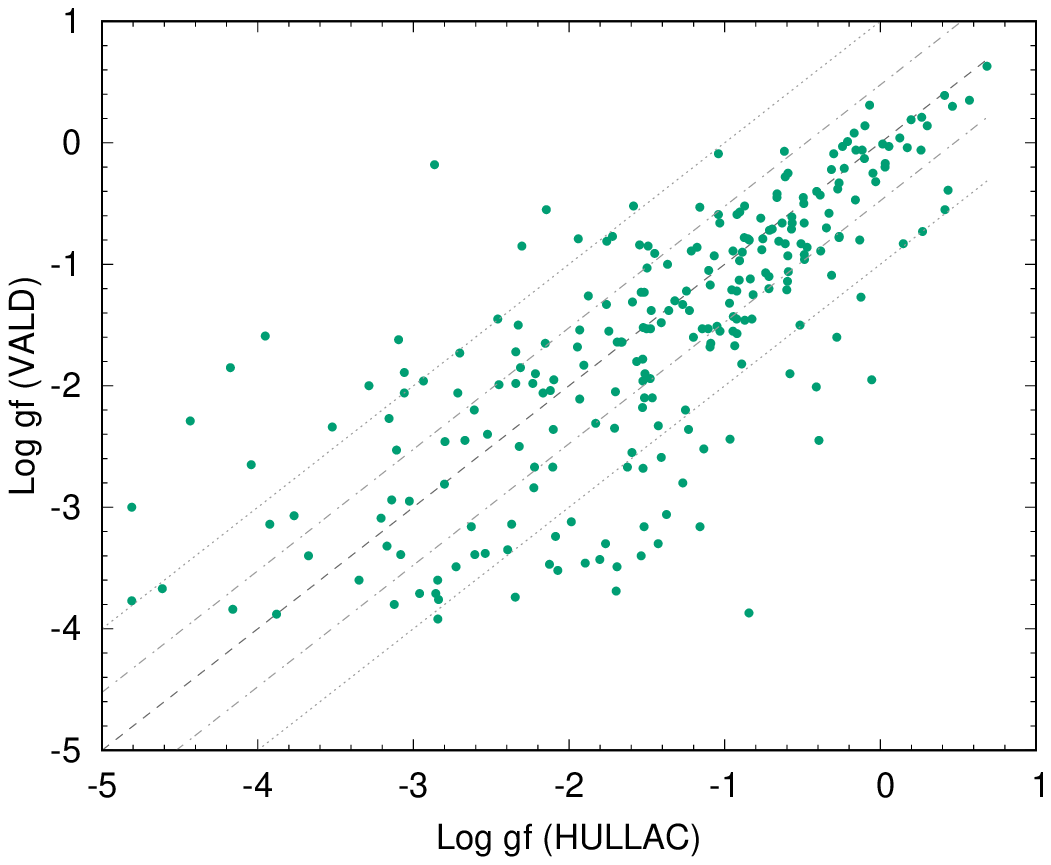}  \\
    \end{tabular}
\caption{
  \label{fig:CeIII}
  Same as Figure \ref{fig:SrII}, but for Ce III.
}
\end{center}
\end{figure*}

\startlongtable
\begin{deluxetable*}{lrrrrrrr}
\tablewidth{0pt}
\tablecaption{Summary of calibrated lines for La III. We list only lines that adopt theoretical $gf$-values with $\lambda>7000$ {\AA} and log $gf > -3$.}
\label{tab:LaIII}
\tablehead{
   &  $\lambda_{\rm vac}$$^a$  &  $\lambda_{\rm air}$$^b$  & Lower level & $E_{\rm lower}$$^c$ & Upper level & $E_{\rm upper}$$^d$ & log $gf^e$ \\
   &  ({\AA})   &  ({\AA})   &   &  (${\rm cm^{-1}}$)  &  &   (${\rm cm^{-1}}$)   & 
}
\startdata
La III  &  13898.270  &  13894.471  &  5$d$ ${\rm ^2D_{3/2}}$ &        0.00  &  4$f$ ${\rm ^2F^o_{5/2}}$ &  7195.14  &  $-$0.749  \\ 
	  &  14100.037  &  14096.183  &  5$d$ ${\rm ^2D_{5/2}}$ &  1603.23  &  4$f$ ${\rm ^2F^o_{7/2}}$ &  8695.41  &  $-$0.587 \\
	  &  17882.977  &  17878.094  &  5$d$ ${\rm ^2D_{5/2}}$ &  1603.23  &  4$f$ ${\rm ^2F^o_{5/2}}$ &  7195.14  &  $-$1.938 \\
\enddata
\tablecomments{
$^a$ Vacuum transition wavelength. \\
$^b$ Air transition wavelength. \\
$^c$ Lower energy level. \\
$^d$ Upper energy level. \\
$^e$ $gf$-value \citep{Tanaka2020}.
}
\end{deluxetable*}
\startlongtable
\begin{deluxetable*}{lrrrrrrr}
\tablewidth{0pt}
\tablecaption{Same as Table \ref{tab:LaIII}, but for Ce III.}
\label{tab:CeIII}
\tablehead{
   &  $\lambda_{\rm vac}$$^a$  &  $\lambda_{\rm air}$$^b$  & Lower level & $E_{\rm lower}$$^c$ & Upper level & $E_{\rm upper}$$^d$ & log $gf^e$ \\
   &  ({\AA})   &  ({\AA})   &   &  (${\rm cm^{-1}}$)  &  &   (${\rm cm^{-1}}$)   & 
}
\startdata
Ce III  &   8100.800  &    8098.573  &  $4f^2$ ${\rm ^1G_4}$      &    7120.000  &  $4f6s$ $\left(\frac{5}{2}\frac{1}{2}\right)_3$  &  19464.460  & $-$2.682 \\ 
          &  11071.353  &  11068.321  &  $4f^2$ ${\rm ^1G_4}$      &    7120.000  &  $4f5d$ ${\rm ^1H^o_5}$  &  16152.320  & $-$2.636 \\ 
          &  11093.386  &  11090.349  &  $4f^2$ ${\rm ^1D_2}$      &  12835.090  &  $4f6s$ $\left(\frac{7}{2}\frac{1}{2}\right)_3$  &  21849.470  & $-$2.870 \\ 
          &  11094.396  &  11091.358  &  $4f5d$ ${\rm ^3F^o_2}$   &    3821.530  &  $4f^2$ ${\rm ^1D_2}$      &  12835.090  & $-$2.669 \\ 
          &  12760.441  &  12756.951  &  $4f^2$ ${\rm ^3H_4}$      &          0.000  &  $4f5d$ ${\rm ^3G^o_4}$  &   7836.720  & $-$1.947 \\ 
          &  12825.133  &  12821.626  &  $4f^2$ ${\rm ^3H_5}$      &    1528.320  &  $4f5d$ ${\rm ^3G^o_5}$  &   9325.510  & $-$1.910 \\ 
          &  12926.644  &  12923.109  &  $4f^2$ ${\rm ^3F_3}$      &    4764.760  &  $4f5d$ ${\rm ^1F^o_3}$   &  12500.720  & $-$2.598 \\ 
          &  13155.108  &  13151.511  &  $4f5d$ ${\rm ^3D^o_1}$  &    8922.050  &  $4f^2$ ${\rm ^3P_1}$      &  16523.660  & $-$2.454 \\ 
          &  13342.833  &  13339.185  &  $4f^2$ ${\rm ^3F_4}$      &    5006.060  &  $4f5d$ ${\rm ^1F^o_3}$  &  12500.720  & $-$0.922 \\ 
          &  13482.540  &  13478.854  &  $4f5d$ ${\rm ^3D^o_2}$  &    9900.490  &  $4f^2$ ${\rm ^3P_2}$      &  17317.490  & $-$2.368 \\ 
          &  13906.349  &  13902.548  &  $4f5d$ ${\rm ^3D^o_3}$  &  10126.530  &  $4f^2$ ${\rm ^3P_2}$      &  17317.490  & $-$2.014 \\ 
          &  13986.034  &  13982.211  &  $4f5d$ ${\rm ^3D^o_1}$  &    8922.050  &  $4f^2$ ${\rm ^3P_0}$      &  16072.040  & $-$2.298 \\ 
          &  14659.167  &  14655.161  &  $4f^2$ ${\rm ^3H_5}$      &    1528.320  &  $4f5d$ ${\rm ^3H^o_6}$  &   8349.990  & $-$2.911 \\ 
          &  15098.510  &  15094.385  &  $4f5d$ ${\rm ^3D^o_2}$  &    9900.490  &   $4f^2$ ${\rm ^3P_1}$     &  16523.660  & $-$2.013 \\ 
          &  15720.131  &  15715.837  &  $4f^2$ ${\rm ^3H_4}$      &          0.000  &  $4f5d$ ${\rm ^3H^o_5}$  &   6361.270  & $-$2.985 \\ 
          &  15851.880  &  15847.550  &  $4f^2$ ${\rm ^3H_5}$      &    1528.320  &  $4f5d$ ${\rm ^3G^o_4}$  &   7836.720  & $-$0.613 \\ 
          &  15961.157  &  15956.797  &  $4f^2$ ${\rm ^3H_4}$      &          0.000  &  $4f5d$ ${\rm ^3G^o_3}$  &   6265.210  & $-$0.721 \\ 
          &  15964.928  &  15960.567  &  $4f5d$ ${\rm ^1D^o_2}$  &    6571.360  &  $4f^2$ ${\rm ^1D_2}$      &  12835.090  & $-$1.272 \\ 
          &  16133.170  &  16128.763  &  $4f^2$ ${\rm ^3H_6}$      &    3127.100  &  $4f5d$ ${\rm ^3G^o_5}$  &   9325.510  & $-$0.509 \\ 
          &  16292.642  &  16288.192  &  $4f^2$ ${\rm ^3F_2}$      &    3762.750  &  $4f5d$ ${\rm ^3D^o_2}$   &   9900.490  & $-$2.050 \\ 
          &  17529.037  &  17524.251  &  $4f5d$ ${\rm ^3P^o_1}$  &  11612.670  &  $4f^2$ ${\rm ^3P_2}$       &  17317.490  & $-$1.873 \\ 
          &  17829.952  &  17825.084  &  $4f^2$ ${\rm ^1D_4}$      &  12835.090  &  $4f5d$ ${\rm ^1P^o_1}$   &  18443.630  & $-$1.546 \\ 
          &  18584.873  &  18579.800  &  $4f^2$ ${\rm ^1G_4}$      &    7120.000  &  $4f5d$ ${\rm ^1F^o_3}$   &  12500.720  & $-$1.850 \\ 
          &  18650.558  &  18645.466  &  $4f^2$ ${\rm ^3F_3}$      &    4764.760  &  $4f5d$ ${\rm ^3D^o_3}$   &  10126.530  & $-$2.173 \\ 
          &  19146.488  &  19141.261  &  $4f^2$ ${\rm ^3H_6}$      &    3127.100  &  $4f5d$ ${\rm ^3H^o_6}$   &   8349.990  & $-$1.496 \\ 
          &  19382.474  &  19377.184  &  $4f^2$ ${\rm ^3F_2}$      &    3762.750  &  $4f5d$ ${\rm ^3D^o_1}$   &   8922.050  & $-$1.373 \\ 
          &  19471.429  &  19466.114  &  $4f^2$ ${\rm ^3F_3}$      &    4764.760  &  $4f5d$ ${\rm ^3D^o_2}$   &   9900.490  & $-$1.210 \\ 
          &  19503.556  &  19498.233  &  $4f^2$ ${\rm ^3H_4}$      &          0.000  &  $4f5d$ ${\rm ^3H^o_4}$   &   5127.270  & $-$1.987 \\ 
          &  19529.457  &  19524.127  &  $4f^2$ ${\rm ^3F_4}$      &    5006.060  &  $4f5d$ ${\rm ^3D^o_3}$   &  10126.530  & $-$2.825 \\ 
          &  20216.315  &  20210.797  &  $4f5d$ ${\rm ^3P^o_0}$  &  11577.160  &  $4f^2$ ${\rm ^3P_1}$       &  16523.660  & $-$1.937 \\ 
          &  20362.493  &  20356.936  &  $4f5d$ ${\rm ^3P^o_1}$  &  11612.670  &  $4f^2$ ${\rm ^3P_1}$       &  16523.660  & $-$2.072 \\ 
          &  20691.296  &  20685.649  &  $4f^2$ ${\rm ^3H_5}$      &    1528.320  &  $4f5d$ ${\rm ^3H^o_5}$   &   6361.270  & $-$1.665 \\ 
          &  20760.800  &  20755.135  &  $4f5d$ ${\rm ^1F^o_3}$  &  12500.720  &  $4f^2$ ${\rm ^3P_2}$       &  17317.490  & $-$2.167 \\ 
          &  21386.074  &  21380.238  &  $4f5d$ ${\rm ^3P^o_2}$  &  12641.550  &  $4f^2$ ${\rm ^3P_2}$       &  17317.490  & $-$1.426 \\ 
          &  22424.692  &  22418.574  &  $4f5d$ ${\rm ^3P^o_1}$  &  11612.670  &  $4f^2$ ${\rm ^3P_0}$       &  16072.040  & $-$2.010 \\ 
          &  23151.096  &  23144.779  &  $4f^2$ ${\rm ^3F_4}$      &    5006.060  &  $4f5d$ ${\rm ^3G^o_5}$   &   9325.510  & $-$2.639 \\ 
          &  25759.188  &  25752.161  &  $4f5d$ ${\rm ^3P^o_2}$  &  12641.550  &  $4f^2$ ${\rm ^3P_1}$       &  16523.660  & $-$1.972 \\ 
          &  26019.036  &  26011.938  &  $4f5d$ ${\rm ^1G^o_4}$  &    3276.660  &  $4f^2$ ${\rm ^1G_4}$       &   7120.000  & $-$1.749 \\ 
\enddata
\tablecomments{
$^a$ Vacuum transition wavelength. \\
$^b$ Air transition wavelength. \\
$^c$ Lower energy level. \\
$^d$ Upper energy level. \\
$^e$ $gf$-value \citep{Tanaka2020}.
}
\end{deluxetable*}

\startlongtable
\begin{deluxetable*}{lrrrrrrr}
\tablewidth{0pt}
\tablecaption{Summary of lines for Th III whose $gf$-values are estimated from the measured line intensities.}
\label{tab:Th3}
\tablehead{
   &  $\lambda_{\rm vac}$$^a$  &  $\lambda_{\rm air}$$^b$  & Lower level & $E_{\rm lower}$$^c$ & Upper level & $E_{\rm upper}$$^d$ & log $gf^e$ \\
   &  ({\AA})   &  ({\AA})   &   &  (${\rm cm^{-1}}$)  &  &   (${\rm cm^{-1}}$)   & 
}
\startdata
Th III  &  10046.6354  &  10043.8823  &  5$f$6$d$ ${\rm ^3H^o_4}$ &          0.000  &  6$d$7$s$ ${\rm ^3D_3}$    &   9953.581  &  $-$1.984  \\ 
	  &  10257.0203  &  10254.2105  &  5$f$6$d$   \ \ \ \ \ \ \            &    6288.221  &  6$d$7$s$ ${\rm ^1D_2}$   & 16037.641  &  $-$1.106 \\
	  &  10260.9778  &  10258.1662  &  5$f$6$d$ ${\rm ^3G^o_4}$ &    8141.749  &  5$f^2$ ${\rm ^3H_5}$         & 17887.409  &       0.979  \\
	  &  10532.8695  &  10529.9844  &  5$f$6$d$ ${\rm ^3G^o_5}$ &  11276.807  &  5$f^2$ ${\rm ^3H_6}$         &  20770.896 &      1.310  \\        
	  &  10581.3571  &  10578.4585  &  5$f$6$d$ ${\rm ^3H^o_6}$ &    8436.826  &  5$f^2$ ${\rm ^3H_5}$         & 17887.409  &  $-$0.994 \\
	  &  10710.5316  &  10703.6448  &  5$f$6$d$ ${\rm ^1H^o_5}$ & 19009.910   &  5$f^2$ ${\rm ^1I_6}$           & 28349.962  &       1.744  \\
          &  11216.3023   &  11213.2319  &  6$d^2$ ${\rm ^3F_4}$        &     6537.817  &  5$f$6$d$ ${\rm ^1F^o_3}$ & 15453.412  &  $-$0.796 \\ 
          &  11227.3116   &  11224.2381  &  5$f$6$d$   \ \ \ \ \ \ \            &     8980.557 &  5$f^2$ ${\rm ^3H_5}$         &  17887.409  &        0.513 \\ 
          &  11428.8103   &  11425.6824  &  7$s^2$ ${\rm ^1S_0}$        &   11961.132 &  5$f$6$d$ ${\rm ^1P^o_1}$ &  20710.949  &  $-$0.485 \\ 
          &  11516.6200   &  11513.4678  &  5$f$6$d$ ${\rm ^3D^o_2}$ &  10180.766  &  5$f^2$ ${\rm ^3F_2}$         &  18863.869  &  $-$0.799 \\ 
          &  11720.8814   &  11717.6743  &  6$d^2$    \ \ \ \ \ \ \               &    4676.432  &  5$f$6$d$ ${\rm ^3P^o_2}$ &  13208.214  &  $-$1.637 \\ 
          &  11810.5436   &  11807.3118  &  6$d^2$ ${\rm ^1G_4}$        &   10542.899 &  5$f$6$d$ ${\rm ^1H^o_5}$ &  19009.910  &       0.657 \\ 
          &  12081.2271   &  12077.9218  &  6$d$7$s$ ${\rm ^3D_2}$   &    7176.107  &  5$f$6$d$ ${\rm ^1F^o_3}$  &  15453.412  &  $-$1.159 \\ 
          &  12320.5004   &  12317.1299  &  5$f$6$d$ ${\rm ^3D^o_1}$ &    7921.088 &  6$d$7$s$ ${\rm ^1D_2}$    &  16037.641  &  $-$0.295 \\ 
          &  12726.1743   &  12722.6938  &  6$d^2$ ${\rm ^3F_2}$        &        63.267  & 5$f$6$d$ ${\rm ^3D^o_1}$  &    7921.088  &  $-$1.209 \\ 
          &  12918.7449   &  12915.2129  &  5$f$6$d$ ${\rm ^3P^o_1}$ &  11123.179  &  5$f^2$ ${\rm ^3F_2}$         &  18863.869  &  $-$0.724 \\ 
          &  13075.4606   &  13071.8856  &  5$f$7$s$   \ \ \ \ \ \ \            &    7500.605  &  5$f^2$ ${\rm ^3H_4}$         &  15148.519  &  $-$0.466 \\ 
          &  13102.2536   &  13098.6709  &  5$f$6$d$ ${\rm ^3P^o_2}$ &  13208.214  &  5$f^2$ ${\rm ^3F_3}$         &  20840.489  &  $-$0.661 \\ 
          &  13445.6702   &  13441.9943  &  6$d^2$ ${\rm ^3F_2}$        &        63.267  &  5$f$7$s$   \ \ \ \ \ \ \             &    7500.605  &  $-$2.137 \\ 
          &  13465.3195   &  13461.6378  &  5$f$7$s$   \ \ \ \ \ \ \            &    2527.095  &  6$d$7$s$ ${\rm ^3D_3}$    &    9953.581  &  $-$1.090 \\ 
          &  13577.6105   &  13573.8997  &  5$f$6$d$ ${\rm ^3F^o_2}$ &      510.758  &  6$d^2$ ${\rm ^3P_1}$        &    7875.824  &  $-$2.619 \\ 
          &  13596.9369   &  13593.2186  &  5$f$6$d$   \ \ \ \ \ \ \            &    3188.301  &  6$d^2$ ${\rm ^1G_4}$        &  10542.899  &  $-$1.960 \\ 
          &  14271.9111    &  14268.0108  &  5$f$6$d$ ${\rm ^3G^o_4}$ &    8141.749  &  5$f^2$ ${\rm ^3H_4}$        &  15148.519  &  $-$0.949 \\ 
          &  14363.1448   &  14359.2200  &  5$f$6$d$ ${\rm ^1H^o_5}$ &  19009.910  &   5$f^2$ ${\rm ^1G_4}$        &  25972.173  &      0.478 \\ 
          &  14766.5150   &  14762.4802  &  5$f$7$s$ ${\rm ^3F^o_2}$  &    3181.502  &  6$d$7$s$ ${\rm ^3D_3}$    &    9953.581  &  $-$2.286 \\ 
          &  14781.3544   &  14777.3154  &  5$f$6$d$   \ \ \ \ \ \ \             &    3188.301  &  6$d$7$s$ ${\rm ^3D_3}$    &    9953.581  &  $-$1.198 \\ 
          &  14958.5620   &  14954.4750  &  6$d^2$ ${\rm ^3F_3}$        &    4056.015  &  5$f$6$d$ ${\rm ^3D^o_3}$  &  10741.150  &  $-$1.640 \\ 
          &  15002.9667   &  14998.8674  &  5$f$6$d$ ${\rm ^3F^o_2}$ &      510.758  &  6$d$7$s$ ${\rm ^3D_2}$     &    7176.107  &  $-$1.319 \\ 
          &  15127.2149   &  15123.0814  &  5$f$6$d$ ${\rm ^3G^o_5}$ &  11276.807  &  5$f^2$ ${\rm ^3H_5}$         &  17887.409  &  $-$0.684 \\ 
          &  15295.6248   &  15291.4461  &  5$f$6$d$ ${\rm ^3H^o_4}$ &          0.000  &  6$d^2$ ${\rm ^3F_4}$        &    6537.817  &  $-$1.523 \\ 
          &  15511.6993   &  15507.4618  &  6$d^2$    \ \ \ \ \ \ \               &    4676.432  &  5$f$6$d$ ${\rm ^3P^o_0}$ &  11123.179  &  $-$1.418 \\ 
          &  16064.3747   &  16059.9860  &  6$d^2$ ${\rm ^3F_2}$        &        63.267  &  5$f$6$d$   \ \ \ \ \ \ \             &    6288.221  &  $-$2.781 \\ 
          &  16212.8109   &  16208.3834  &  5$f$6$d$   \ \ \ \ \ \ \             &    8980.557 &  5$f^2$ ${\rm ^3H_4}$         &  15148.519  &  $-$1.665 \\ 
          &  16327.1959   &  16322.7369  &  6$d^2$ ${\rm ^3F_3}$        &    4056.015  &  5$f$6$d$ ${\rm ^3D^o_2}$ &  10180.766  &  $-$1.239 \\ 
          &  16488.8131   &  16484.3100  &  6$d^2$    \ \ \ \ \ \ \               &    4676.432  &  5$f$6$d$ ${\rm ^3D^o_3}$ &  10741.150  &  $-$0.384 \\ 
          &  16577.9550   &  16573.4271  &  6$d$7$s$ ${\rm ^3D_2}$    &    7176.107  &  5$f$6$d$ ${\rm ^3P^o_2}$ &  13208.214  &  $-$1.372 \\ 
          &  17073.9505   &  17069.2888  &  5$f$6$d$ ${\rm ^3D^o_2}$ &  10180.766  &  6$d$7$s$ ${\rm ^1D_2}$   &  16037.641  &  $-$1.313 \\ 
          &  17494.5296   &  17489.7524  &  5$f$6$d$   \ \ \ \ \ \ \            &    4826.826  &  6$d^2$ ${\rm ^1G_4}$        &  10542.899  &  $-$0.741 \\ 
          &  17517.0190   &  17512.2353  &  6$d$7$s$ ${\rm ^3D_1}$    &    5523.881  &  5$f$6$d$ ${\rm ^3P^o_0}$ &  11232.615  &  $-$1.441 \\ 
          &  17814.4804   &  17809.6160  &  5$f$6$d$   \ \ \ \ \ \ \            &    4826.826  &  6$d^2$ ${\rm ^3P_2}$        &  10440.237  &  $-$2.318 \\ 
          &  17859.3810   &  17854.5071  &  6$d$7$s$ ${\rm ^3D_1}$    &    5523.881  &  5$f$6$d$ ${\rm ^3P^o_1}$ &  11123.179  & $-$1.467 \\ 
          &  18182.3784   &  18177.4150  &  6$d$7$s$ ${\rm ^3D_3}$    &    9953.581  &  5$f$6$d$ ${\rm ^1F^o_3}$ &  15453.412  &  $-$2.372 \\ 
          &  18240.3369   &  18235.3613  &  5$f$6$d$ ${\rm ^3G^o_3}$ &    5060.544  &  6$d^2$ ${\rm ^1G_4}$        &  10542.899  &  $-$1.492 \\ 
          &  18588.4201   &  18583.3460  &  5$f$6$d$ ${\rm ^3G^o_3}$ &    5060.544  &  6$d^2$ ${\rm ^3P_2}$        &  10440.237  &  $-$2.281 \\ 
          &  18880.4240   &  18875.2700  &  5$f$6$d$ ${\rm ^3D^o_3}$ &  10741.150  &  6$d$7$s$ ${\rm ^1D_2}$    &  16037.641  &  $-$1.312 \\ 
          &  19947.4412   &  19941.9969  &  6$d^2$ ${\rm ^3P_2}$        &  10440.237  &  5$f$6$d$ ${\rm ^1F^o_3}$  &  15453.412  &  $-$0.178 \\ 
          &  19947.6467   &  19942.2025  &  5$f$6$d$ ${\rm ^3F^o_2}$ &      510.758  &  6$d$7$s$ ${\rm ^3D_1}$    &    5523.881  &  $-$1.086 \\ 
          &  20010.8977   &  20005.4362  &  6$d^2$ ${\rm ^3F_2}$        &        63.267  &  5$f$6$d$ ${\rm ^3G^o_3}$ &    5060.544  &  $-$1.190 \\ 
          &  20306.4569   &  20300.9148  &  6$d^2$ ${\rm ^3F_3}$        &    4056.015  &  5$f$6$d$   \ \ \ \ \ \ \            &    8980.557  &  $-$0.442 \\ 
          &  20364.4720   &  20358.9157  &  6$d^2$ ${\rm ^1G_4}$        &  10542.899  &  5$f$6$d$ ${\rm ^1F^o_3}$ &  15453.412  &  $-$0.274 \\ 
          &  20437.2044   &  20431.6273  &  5$f$6$d$ ${\rm ^3G^o_3}$ &    5060.544  &  6$d$7$s$ ${\rm ^3D_3}$  &    9953.581  &  $-$1.782 \\ 
          &  20992.7055   &  20986.9762  &  6$d^2$ ${\rm ^3F_2}$        &        63.267  &  5$f$6$d$   \ \ \ \ \ \ \            &    4826.826  &  $-$0.743 \\ 
          &  21101.5437   &  21095.7851  &  6$d^2$ ${\rm ^3F_4}$         &    6537.817 &  5$f$6$d$ ${\rm ^3G^o_5}$ &  11276.807  &        0.314 \\ 
          &  21398.1215   &  21392.2817  &  6$d$7$s$ ${\rm ^1D_2}$    &  16037.641 &  5$f$6$d$ ${\rm ^1P^o_1}$  &  20710.949  &  $-$0.251 \\ 
          &  21473.5815   &  21467.7282  &  6$d$7$s$ ${\rm ^3D_1}$    &    5523.881  &  5$f$6$d$ ${\rm ^3D^o_2}$ &  10180.766  &  $-$2.887 \\ 
          &  21509.9507   &  21504.0820  &  5$f$7$s$   \ \ \ \ \ \ \             &    2527.095  &  6$d$7$s$ ${\rm ^3D_2}$    &    7176.107  &  $-$0.982 \\ 
          &  22689.2716   &  22683.0812  &  5$f$6$d$ ${\rm ^3D^o_3}$ &  10741.150  &  5$f^2$ ${\rm ^3H_4}$         &  15148.519  &  $-$1.177 \\ 
          &  23628.9815   &  23622.5344  &  5$f$7$s$ ${\rm ^3F^o_4}$  &    6310.808  &  6$d^2$ ${\rm ^1G_4}$       &  10542.899  &  $-$1.267 \\ 
          &  23790.6455   &  23784.1542  &  6$d^2$ ${\rm ^3F_4}$         &    6537.817  &  5$f$6$d$ ${\rm ^3D^o_3}$ &  10741.150  &  $-$0.732 \\ 
          &  24005.7196   &  23999.1688  &  5$f$6$d$ ${\rm ^3F^o_2}$ &      510.758  &  6$d^2$    \ \ \ \ \ \ \               &    4676.432  &  $-$1.437 \\ 
          &  24475.4075   &  24468.7299  &  6$d^2$ ${\rm ^3F_3}$        &    4056.015  & 5$f$6$d$ ${\rm ^3G^o_4}$  &    8141.749  &       0.013 \\ 
          &  25335.2327   &  25328.3228  &  6$d$7$s$ ${\rm ^3D_2}$    &    7176.107  &  5$f$6$d$ ${\rm ^3P^o_1}$ &  11123.179  &  $-$1.584 \\ 
          &  25897.4091   &  25890.3447  &  5$f^2$ ${\rm ^3H_4}$         &  15148.519  &  5$f$6$d$ ${\rm ^1H^o_5}$ &  19009.910  &  $-$0.921 \\ 
          &  27282.4514   &  27275.0145  &  5$f$6$d$   \ \ \ \ \ \ \            &    6288.221  &  6$d$7$s$ ${\rm ^3D_3}$     &    9953.581  &  $-$2.707 \\ 
          &  27451.6126   &  27444.1264  &  5$f$7$s$ ${\rm ^3F^o_4}$ &    6310.808  &  6$d$7$s$ ${\rm ^3D_3}$     &    9953.581  &  $-$0.556 \\ 
          &  28050.1492   &  28042.4984  &  6$d$7$s$ ${\rm ^3D_2}$    &    7176.107 &  5$f$6$d$ ${\rm ^3D^o_3}$  &  10741.150  &  $-$1.066 \\ 
          &  28206.6978   &  28199.0038  &  5$f$6$d$ ${\rm ^3F^o_2}$ &      510.758  &  6$d^2$ ${\rm ^3F_3}$         &    4056.015  &  $-$1.924 \\ 
          &  29790.3617   &  29782.2484  &  6$d^2$ ${\rm ^3P_1}$        &    7875.824  &  5$f$6$d$ ${\rm ^3P^o_0}$  &  11232.615  &  $-$2.471 \\ 
          &  29855.0580   &  29846.9168  &  5$f$6$d$   \ \ \ \ \ \ \            &    3188.301  &  6$d^2$ ${\rm ^3F_4}$         &    6537.817  &  $-$1.015 \\ 
\enddata
\tablecomments{
$^a$ Vacuum transition wavelength. \\
$^b$ Air transition wavelength. \\
$^c$ Lower energy level. \\
$^d$ Upper energy level. \\
$^e$ $gf$-value estimated in Section \ref{sec:actinide}.
}
\end{deluxetable*}

\bibliography{reference}
\bibliographystyle{aasjournal}

\end{document}